\documentclass{article}
\usepackage{natbib}
\usepackage{ulem}
\usepackage{physics}
\usepackage{gensymb}
\usepackage{threeparttable}

\usepackage{latexsym, color}
\usepackage{amsmath}
\usepackage{mathtools}
\usepackage{amsfonts}
\usepackage{amssymb}
\usepackage{psfrag}
\usepackage{graphicx}
\usepackage{epsfig}
\usepackage{multirow}
\usepackage{multicol}
\usepackage{subfigure}
\usepackage{hyperref}
\usepackage{cleveref}
\usepackage{qtree} 
\usepackage{multicol} 
\usepackage{tikz} 
\usepackage{forest} 
\usepackage{color}
\usepackage{calc}
\usepackage{pdfpages}
\usepackage{booktabs}
\usepackage{float}
\usepackage[TS1, T1]{fontenc}
\usepackage[utf8]{inputenc}

\renewcommand{\bibsection}{\section*{References}}

\newtheorem{definition}{Definition}
\newtheorem{remark}{Remark}

\usepackage{xurl}

\begin{document}

\title{Predicting Dengue Outbreaks: A Dynamic Approach with Variable Length Markov Chains and Exogenous Factors}

\author{Marília Gabriela Rocha, Nancy L. Garcia\thanks{Statistics Department, Universidade Estadual de Campinas(UNICAMP), Brazil}}



\maketitle

\begin{abstract}

{Variable Length Markov Chains with Exogenous Covariates (VLMCX) are stochastic models that use Generalized Linear Models to compute transition probabilities, taking into account both the state history and time-dependent exogenous covariates. The beta-context algorithm selects a relevant finite suffix (\textit{context}) for predicting the next symbol. This algorithm estimates flexible tree-structured models by aggregating irrelevant states in the process history and enables the model to incorporate exogenous covariates over time.

This research uses data from multiple sources to extend the beta-context algorithm to incorporate time-dependent and time-invariant exogenous covariates. Within this approach, we have a distinct Markov chain for every data source, allowing for a comprehensive understanding of the process behavior across multiple situations, such as different geographic locations. Despite using data from different sources, we assume that all sources are independent and share identical parameters - we explore contexts within each data source and combine them to compute transition probabilities, deriving a unified tree. This approach eliminates the necessity for spatial-dependent structural considerations within the model. Furthermore, we incorporate modifications in the estimation procedure to address contexts that appear with low frequency.

Our motivation was to investigate the impact of previous dengue rates, weather conditions, and socioeconomic factors on subsequent dengue rates across various municipalities in Brazil, providing insights into dengue transmission dynamics.}

\end{abstract}

\renewcommand\thefootnote{}
\footnotetext{\textbf{Abbreviations:} VLMC, Variable Length Markov Chain; VLMCX, Variable Length Markov Chains with Exogenous Covariates; WHO, World Health Organization.}

\renewcommand*{\thefootnote}{\arabic{footnote}}
\setcounter{footnote}{1}

\section{Dengue: an overview of the disease and its global impact}\label{sec:dengue}

Dengue is an urban arboviral disease, meaning it is caused by viruses transmitted by mosquitoes. The most common arboviral diseases in urban environments are Dengue, Chikungunya, and Zika, transmitted by infected female mosquitoes of the genus \textit{Aedes}, primarily \textit{Aedes aegypti}. These mosquitoes can be infected with four different serotypes of the Dengue virus: DENV-1, DENV-2, DENV-3, and DENV-4. Once infected, a female mosquito becomes a permanent disease vector, with a 30 to 40\% chance of transmitting the virus to its offspring \citep{rita2016}. These mosquitoes, known for their strictly urban behavior, lay eggs just above clean water surfaces in containers like cans, bottles, and tires, which hatch within minutes upon contact with rising water levels. On that account, the mosquito density is higher in the summer, when the periods of elevated rainfall increase the number of breeding sites. Additionally, higher temperatures during the summer accelerate the mosquito's development through the egg-larva-adult stages. 

Since dengue is primarily transmitted amongst humans through mosquito vectors, preventing the spread of \textit{Aedes aegypti} is essential. This is most effective during the insect aquatic phase, focusing on larvae and pupa removal or covering potential breeding sites \citep{rita2016}. As such, public awareness and continuous home monitoring are crucial stances to keep the disease rates under control, and both are heavily dependent on forethought public policies and institutional actions designed to create an effective sanitary culture.

Regarding global impact, dengue's incidence has grown dramatically in recent decades. According to the World Health Organization (WHO), before the COVID-19 pandemic in 2020, dengue fever was ranked amongst the top ten global health threats \footnote{Available at \url{https://www.who.int/news-room/spotlight/ten-threats-to-global-health-in-2019}. Accessed July 28, 2024.}. The disease is now endemic in over 100 countries and the most severely affected regions are the Americas, Southeast Asia, and the Western Pacific, with Asia contributing to approximately 70\% of the global disease burden. Notably, dengue is extending its reach to new areas, including Europe, leading to explosive outbreaks. In 2010, local transmission was reported for the first time in France and Croatia, and imported cases were detected in three other European countries \footnote{Available at \url{https://www.who.int/news-room/fact-sheets/detail/dengue-and-severe-dengue}. Accessed July 28, 2024.}.

The incidence of dengue in Brazil follows endemic and epidemic cycles, with explosive outbreaks occurring approximately every 4 to 5 years. Since the introduction of the virus in the country (1981), more than seven million cases have been reported. In the last two years (2022 and 2023), Brazil faced its highest recorded dengue-related deaths: at least 1,016 \citep{brasil2022monitoramento} and 1,179 \footnote{Available at \url{https://www.gov.br/saude/pt-br/assuntos/saude-de-a-a-z/a/aedes-aegypti/monitoramento-das-arboviroses}. Accessed July 28, 2024.} fatalities, respectively, as reported by the Notifiable Diseases Information System (Sistema de Informa\c c\~ao de Agravos de Notifica\c c\~ao - SINAN) \footnote{Available at \url{http://portalsinan.saude.gov.br/dengue}. Accessed August 19, 2024.}. Climate change has allowed the disease vector to adapt and spread to regions where it was not previously common \footnote{Available at \url{https://butantan.gov.br/noticias/aumento-historico-de-temperatura-leva-a-disseminacao-da-dengue-em-todo-o-brasil}. Accessed July 28, 2024.}. There was an explosion of cases in 2024 - on July 29, 2024, there were 6,425,311 cases and 4,856 confirmed deaths \footnote{Available at \url{https://www.gov.br/saude/pt-br/assuntos/saude-de-a-a-z/a/aedes-aegypti/monitoramento-das-arboviroses}. Accessed July 28, 2024.}. According to specialists, the excessive heat waves and rainfall and the simultaneous circulation of all four serotypes along with the growth of urban population are among the main reasons for the explosion of the disease.

According to the National Institute of Meteorology (Instituto Nacional de Metereologia - INMET), temperatures in Brazil have consistently exceeded historical averages since the 1990s. This warming climate provides favorable conditions for the adaptation and proliferation of the dengue-transmitting mosquito. In the southern part of the country, increased rainfall and higher average temperatures led the region to be ranked second in 2022's dengue incidence rates, according to the Epidemiological Bulletin of the Ministry of Health \footnote{All Epidemiological Bulletin of the Ministry of Health are available at \url{https://www.gov.br/saude/pt-br/centrais-de-conteudo/publicacoes/boletins/epidemiologicos/edicoes/2023}. Accessed August 19, 2024.} - until mid-2015, the presence of dengue in the region was variable and not very significant, as shown by \citet{dalla2018panorama}.



\section{Reviewing forecasting models for dengue incidence}\label{sec:modeldenguereview}

In addition to efforts in dengue vaccine development and combating \textit{Aedes aegypti}, another strategy for dengue-endemic countries is the development of forecasting models for outbreak prediction - this way, public health systems can be prepared in terms of resources and protocols for the high influx of patients. WHO supports this strategy by providing financial assistance for innovative surveillance systems, enhancing prevention, control, and forecasting efficiency \citep{world2012global}. \citet{stanaway2016global} conducted a literature review, summarizing researchers' efforts to gather and analyze data, improving understanding of the relational factors influencing disease spread. The review also emphasizes the evolution of various predictive modeling methods, including statistical and mathematical analyses as well as machine learning techniques.

In this section, we will rely on the literature review conducted by \citet{stanaway2016global} to offer a brief overview of common relational factors used by researchers in dengue forecast modeling and the predictive modeling methods they employ.

As pointed out by \citet{stanaway2016global}, derived from their literature review on dengue forecast modeling, there are direct and indirect factors that can be correlated to the number of dengue cases. Direct factors are the ones that directly affect the mosquito's life cycle such as climate factors, mosquito density, dengue virus serotypes, and bite rate. Indirect factors are those that do not have a direct impact on the number of mosquito larvae but may be associated with the occurrence of a dengue epidemic such as geography, spatial interaction, population movement, immunology, and environment. 

In our study, we included some of the factors that have public available information easily accessible such as temperature and precipitation along with fixed attributes of municipalities, such as poverty rates and urban population percentages. Of these factors, rainfall significantly impacts the incubation period of mosquitoes, as they require still or standing water to complete their life cycle \citep{buczak2012data}. This way, unusual weather conditions, such as drought and higher temperatures associated with phenomena like \textit{El Ni\~no}, adversely affect mosquito breeding habitats and populations. Interestingly, some studies suggest that excessive rainfall can lead to a decline in dengue epidemics. This may be attributed to the disruptive effects of fast-moving water flows on larvae survival and growth, as well as a reduction in the larval population \citep{hsu2013analysis, thammapalo2005climatic, arcari2007regional}.  

There are many works in the literature that also study the need to incorporate spatial analysis into the modeling of outbreaks of dengue, especially in terms of prediction. For example, \citet{yu2014online} proposed a spatial-temporal model incorporating population density, environmental conditions, and infrastructure factors. However, \citet{c2015surveillance} argued that relying solely on spatial information may not be sufficient, as \textit{Ae. aegypti} is strongly influenced by local climate triggers. \citet{thiruchelvam2018correlation} found that dengue incidences were localized, and feedback models tailored to specific regions did not benefit from data from neighboring areas.

The majority of dengue outbreaks are observed in communities with unhygienic housing conditions and malnourished populations, contributing to lower immunity and an increased risk of disease infection \citep{world2014global}. Additionally, urban growth creates numerous breeding areas for dengue vectors - mosquitoes adapt well to this environment as they have a preference for feeding on humans over animals. Moreover, the ability of mosquitoes to fly within a range of 100-500 meters to find food and breeding sites increases the risk of mosquito-borne infections in urban areas, as houses fall within this flight range.

In their survey, \citet{stanaway2016global} identified common methods employed for detecting dengue outbreaks, forecasting future dengue cases, and determining critical regions. They reported a total of 966 models created for the analysis of dengue epidemics, with 545 using regression methods, 220 using temporal series, 76 using neural networks, 50 using decision trees, 23 using support vector machine, 20 using k-means, 17 using association rules, 9 using lattice models and 6 using K-Nearest Neighbor. However, an important limitation of some of the existing works, especially the ones relying on machine learning methods, is that they provide little or no explanations for the predictions, especially in the case of more complex models. Also, it is difficult to compare the efficiency of different methods since they use distinct metrics, and different datasets, often unavailable or hard to obtain. Addressing these limitations, \citet{aleixo2022predicting} utilized a so-called explainable boosted decision tree model - CatBoost \citep{prokhorenkova2018catboost} for dengue outbreak detection in Rio de Janeiro, Brazil. This approach allows practitioners to comprehend how the model uses available information for predictions. Importantly, not only did they make all data and code publicly available, but also evaluated the model using multiple error metrics for regression and classification, and a detailed analysis per district, month of the year, and prediction span.

Aligned to provide an explainable and comparable model for dengue outbreak detection, we propose to use a model-based methodology, Variable Length Markov Chains with Exogenous Covariates (VLMCX) and the beta-context algorithm, introduced by \citet{zanin2022variable}, integrate both time-dependent covariates - temperature and precipitation levels, along with time-invariant covariates - poverty rates and urban population percentages, allowing us to investigate the complex dynamics of dengue transmission. The model demonstrated satisfactory performance in predicting dengue incidence, especially for high-incidence cases, showing its potential for early detection and proactive management of outbreaks.

\section{Context Algorithm}\label{sec:context_algorithm}

Stochastic chains with memory of variable length constitute an interesting family of stochastic chains of infinite order on a finite alphabet. The idea is that for each past, only a finite suffix of the past, called {\sl context}, is enough to predict the next symbol. These models were first introduced in the information theory literature by \citet{rissanen1983universal} as a universal tool to perform data compression. It focuses on capturing \textit{contexts}, which are relevant subsets of the past string, along with the corresponding counts of conditional occurrences, to
estimate the transition probabilities to the next symbol using maximum likelihood estimation. To do this, he proposed the Context Algorithm - a tool capable of efficiently compressing long strings without prior source knowledge. This algorithm was subsequently improved in the works of \citet{rissanen1986complexity} and \citet{furlan1990contribution}.  In general, better data compression results from larger contexts, which increases both their quantity and the model's complexity. However, adding each new context carries a cost that must be balanced against the additional compression gain it provides. Consequently, with a reasonable selection of the parameter, the model's complexity does not exceed that of the original data source. Since all methods related to stochastic chains with variable-length memory rely on the Context Algorithm, we explore it in detail in the Supplementary material. When the order of the tree is bounded, irrespective of the sample size, the Context Algorithm consistently estimates both the context length and the corresponding transition probabilities.

Markov chains with variable length were further refined from a statistical perspective by \citet{buhlmann1999variable}, which introduced the name Variable Length Markov Chains (VLMC). They demonstrated that the context algorithm remains consistent even when the order of the chain is permitted to expand with the sample size. VLMCs have been applied to many fields such as proteomics \citep{flesia2009unsupervised}, genomics \citep{dalevi2006bayesian}, linguistics \citep{galves2012context}, music \citep{marchini2010unsupervised}, and much more. However, in all these applications, no covariates were used to improve the fit of the model.  Based on the context algorithm, \citet{zanin2022variable} developed a backward selection algorithm for Markov chains with variable length incorporating exogenous covariates, considering not only the time sequence of symbols of response states but also a time sequence of covariates that are associated with parameters that define the transition probability in terms of logistic regression - the transition probability of the VLMC depends on both past states and past values of covariates. They call their algorithm VLMCX and prove consistency under the hypothesis that the tree's order is bounded, irrespective of the sample size. 

In this work, to model the dengue data, we propose some modifications to incorporate samples from different locations leading to not only time-dependent covariates being available but also location-specific covariates such as poverty rates and concentration of urban population. Also, the VLMCX algorithm is based on asymptotic results (such as the likelihood ratio test) which were not always the case in our dataset. Thus, some modifications were introduced in the algorithm to apply them to cases where the observed frequency of a context was small. 

\subsection{Notation and definitions}

Before exploring the description of VLMC and VMLCX and the modified version proposed in this study, it is essential to establish some fundamental definitions. These definitions are essentially the same as used in \citet{zanin2022variable}, but have a few modifications to consider different sources and time-invariant covariates. The definitions for VLMC and VLMCX are specific cases derived from these broader scenarios. 

Let ${\cal Y} = \{1, ..., p\}$ represent the finite alphabet and ${\cal S} = \{1, ..., q\}$ denote the finite set of potential sources (in the application, as we aim to predict dengue rates in Brazil, the sources refer to different geographic locations, specifically municipalities). For each independent source, $s \in{\cal S}$, consider a stochastic process $(Y_t(s))_{t \in \mathbb{Z}}$ with values in the alphabet  ${\cal Y}$. Denote by $y_i^j = y_j, y_{j-1}, ..., y_i \ (i < j, i, j \in \mathbb{Z} \cup \{-\infty, \infty\}, y_j \in {\cal Y})$ the string that represents the states visited by the process from time $i$ to time $j$. Notice that it is written in reverse order. 

Suppose that the transition probabilities of the process depend on the previous states through a set of parameters, $d$ time-varying exogenous covariates and $m$ time-invariant exogenous covariates (fixed characteristics). For each source, $s \in{\cal S}$, denote the $b$-th time-varying covariate value at time $t$ by $x_{tb}(s)$, $b = 1, ..., d$, and let $\mathbf{x}_t(s) = (x_{t1}(s), ..., x_{td}(s))$. Also, denote the $v$-th time-invariant covariate value by $z_v(s)$, $v = 1, ..., m$, and let $\mathbf{z}(s) = (z_1(s), ..., z_m(s))$.

Further, define 

\begin{align*}
    \mathbf{x}_i^j(s) = (1, \mathbf{x}_j(s), \mathbf{x}_{j-1}(s), ..., \mathbf{x}_i(s))
\end{align*}

as the vector of time-varying covariates for source $s$ from time $i$ to time $j$. To estimate the intercept of the regression we need to include the number one. Notice that we are considering all the covariates to be deterministic.

The proposed model (given by \eqref{prob_modified_vlmcx})  writes the transition probabilities into the $p$ states by a multinomial linear regression with parameters that can depend on the previous history.


\begin{definition}[Context]
\label{def_context_modified_vlmcx}
    For each independent source, $s \in{\cal S}$, let $(Y_t(s))_{t \in \mathbb{Z}}$ be a stationary process with values $Y_t(s) \in {\cal Y}$, $(\mathbf{x}_t(s))_{t \in \mathbb{Z}}$ a d-dimensional vector of deterministic time-varying exogenous covariates and $(\mathbf{z}(s))$ a m-dimensional vector of deterministic time-invariant exogenous covariates, both in a compact set. Denote by $c: {\cal Y}^\infty \rightarrow \bigcup_{l = 0}^{\infty} {\cal Y}^l$ a (projection) function which maps $c: y_{-\infty}^0 \rightarrow y_{-l+1}^0$, where $l$ is defined by 
    \begin{align*} 
        l = l(y_{-\infty}^0) &= \min\{k; \prod_{s=1}^q P[Y_1(s) = y_1 \mid Y_{- \infty}^0(s) = y_{- \infty}^0, \mathbf{x}_{-\infty}^0(s), \mathbf{z}(s)] \\
        &\quad \quad \quad \quad = \prod_{s=1}^q P[Y_1(s) = y_1 \mid Y_{-k+1}^0(s) = y_{-k+1}^0, \mathbf{x}_{-k+1}^0(s), \mathbf{z}(s)] \\
        &\quad \quad \quad \quad \quad \text{for all} \ \mathbf{x}_{-\infty}^0(s), \mathbf{z}(s) 
        \ \mbox{and for all} \ y_1 \in {\cal Y}\},
    \end{align*} 
    where $l \equiv 0$ corresponds to independence.

    Letting $u := y_{-l+1}^0$ and $\pi_j(s):= P_{\boldsymbol{\theta}}(Y_1(s) = j \mid Y_{-l+1}^0(s) = u, \mathbf{x}_{-l+1}^0(s), \mathbf{z}(s))$, where $\mathbf{x}_{-l+1}^0(s)$ and $\mathbf{z}(s)$ will represent the values of the covariates at source $s$,

    \begin{align}\label{prob_modified_vlmcx}
        \boldsymbol{\pi}(s) = \mathbf{g}(\mathbb{X}^{\top}(s)\boldsymbol{\theta}^u)
    \end{align}

    where 
  
    \begin{align*}
        \mathbb{X}(s) = 
        \begin{bmatrix} 
            (\mathbf{x}_{-l+1}^0(s), \mathbf{z}(s))^{\top} & 0 & \ldots & 0 \\
            0 & (\mathbf{x}_{-l+1}^0(s), \mathbf{z}(s))^{\top} & \ldots & 0 \\
            \vdots & \vdots & \ddots & \vdots \\
            0 & 0 & \ldots & (\mathbf{x}_{-l+1}^0(s), \mathbf{z}(s))^{\top}
        \end{bmatrix},
    \end{align*}

    \begin{align*}
        \boldsymbol{\theta}^u = 
        \begin{bmatrix} 
            \boldsymbol{\theta}_1^u \\ \vdots \\ \boldsymbol{\theta}_p^u
        \end{bmatrix}, \mbox{ and }
        \boldsymbol{\pi}(s) =
        \begin{bmatrix}
            \pi_1(s) \\ \vdots \\ \pi_p(s)
        \end{bmatrix}.
    \end{align*} 
     
    Note that $\mathbb{X}(s)$ is a $(1+dl+m)p$ by $p$ matrix and $\boldsymbol{\theta}^u$ is a $(1+dl+m)p$ dimensional vector with components

    \begin{align*}
        \boldsymbol{\theta}_j^u = (\alpha_j^u, \boldsymbol{\beta}_{j, 0}^u, ..., \boldsymbol{\beta}_{j, (-l+1)}^u, \boldsymbol{\gamma}_j^u)^{\top}
    \end{align*}
    
    corresponding to the vector of coefficients associated with the past states $u = y_{-l+1}^0$ for transitioning into state $j \in {\cal Y}$, where $\boldsymbol{\beta}_{j, t}^u = (\beta_{j, t1}^u, ..., \beta_{j, td}^u)^{\top}$ is the vector of coefficients corresponding to the $d$ time-varying covariates at time $t = 0, ..., -l+1$ and $\boldsymbol{\gamma}_j^u = (\gamma_{j, 1}^u, ..., \gamma_{j, m}^u)^{\top}$ is the vector of coefficients corresponding to the $m$ time-invariant covariates. The link function $\mathbf{g}$ is a one-to-one mapping from a p-dimensional region $D \subset \mathbb{R}^{p}$ to the simplex $\{(\pi_1, ..., \pi_p)^{\top}, \pi_j > 0, \sum \pi_j = 1\}$.

    Then $c(\cdot)$ is called the \textit{beta-context} function, and $c(y_{- \infty}^{0})$ is called the \textit{beta-context} for the transition at time 1 with associated parameter vector $\boldsymbol{\theta}^u$.
\end{definition}

\begin{remark}
    If the link function $\mathbf{g}$ is chosen to be multinomial logistic (softmax function), then
    \begin{align*}
        \pi_j & := P_{\boldsymbol{\theta}}(Y_1 = j \mid Y_{-l+1}^0 = u, \mathbf{x}_{-l+1}^0, \mathbf{z}) \\
        & = \frac{\exp(\alpha_j^u + \sum_{t=-l+1}^0 \sum_{b=1}^d x_{tb}\beta_{j,tb}^u + \sum_{v = 1}^m z_v\gamma_{j, v}^u)}{\sum_{i=1}^p \exp(\alpha_i^u + \sum_{t=-l+1}^0 \sum_{b=1}^d x_{tb}\beta_{i, tb}^u + \sum_{v = 1}^m z_r\gamma_{j, v}^u)} \quad \text{for all} \quad j\in {\cal Y}.
    \end{align*}
\end{remark}

\begin{definition}[Order of the time-varying covariate parameters]  \label{def_beta_length}

    Consider a parameter vector $\boldsymbol{\beta}_j^u$, $j = 1, ..., p$ associated with a beta-context $u$. The length of $\boldsymbol{\beta}_j^u$, which represents the number of steps where covariate values have a significant contribution to the model, is defined as
    
    \begin{align*}
        h := h_u = \left|{\boldsymbol{\beta}_j^u}\right| = 1 - \underset{k = 0, ..., -l+1}{\min} \{k: \boldsymbol{\beta}_{j,k}^u \neq 0\}, j = 1, ..., p.
    \end{align*}
    
    If $\boldsymbol{\beta}_{j, 0}^u = ... = \boldsymbol{\beta}_{j, (-l+1)}^u = 0$, then  $\left|{\boldsymbol{\beta}_{j}^u}\right| = 0$.
\end{definition}

The proposed model accommodates situations where $h \leq l$, including cases where $h < l$, that is, the transition probability may rely on a more extended history of state transitions and time-invariant covariates while considering only the more recent history of time-varying covariates. 

\begin{definition}[Order of the state transitions and time-invariant covariates]
\label{def_order_modified_vlmcx}
    For each independent source, $s \in{\cal S}$, let $(Y_t(s))_{t \in \mathbb{Z}}$, $(\mathbf{x}_t(s))_{t \in \mathbb{Z}}$, $\mathbf{z}(s)$, $c(\cdot)$ and $l(\cdot)$ be defined as in Definition \ref{def_context_modified_vlmcx}. Let $0 \leq \eta < \infty$ be the smallest integer such that

    \begin{align*}
        |c(y_{-\infty}^0)| = l(y_{-\infty}^0) \leq \eta, \quad \text{for all} \quad y_{-\infty}^0 \in {\cal Y}^{\infty}.
    \end{align*}
    
    Then $c(\cdot)$ is called a beta-context function of order $\eta$ and we have a beta-context model of order $\eta$. 
\end{definition}

\begin{definition}[Beta-context tree]
\label{def_tree_modified_vlmcx}

    Let $c(\cdot)$ be a beta-context function of a beta-context model of order $\eta$. The $\left|{\cal Y}\right|$-ary beta-context rooted tree $\tau$ is defined as

    \begin{align*}
        \tau := \tau_c = \{u : u = c(y_{-\eta+1}^0), y_{-\eta+1}^0 \in {\cal Y}^{\eta} \}
    \end{align*}
    
    with an associated parameter tree

    \begin{align*}
        \tau_{\theta} = \{(u, \boldsymbol{\theta}^u): u \in \tau\}
    \end{align*}
    
    where $\boldsymbol{\theta}^u$ is defined in Definition \ref{def_context_modified_vlmcx}.
\end{definition}


Besides the definitions above, to understand the statistical perspective of the context algorithm and the beta-context algorithm, it is essential to have a clear understanding of the terms siblings, children, and parents used in this text, which are defined as follows.

\begin{definition}[Siblings, children and parents]
\label{def_siblings}
    Let $u_1 = uw \in \tau$ and $u_2 = uw' \in \tau$, for $w$, $w' \in {\cal Y}$ and $u \in {\cal Y}^{\infty}$, be two contexts differing only by the last nodes. Then $u_1$ and $u_2$ are called siblings in $\tau$ and this relationship is denoted by $u_1 \wr u_2$. In addition, $u$ is called the parent of $u_1$ and $u_2$, and $u_1$ and $u_2$ are called the children of $u$.
\end{definition}

In this section, we introduce modifications to the beta-context algorithm to accommodate time-dependent and time-invariant exogenous covariates, considering data collected from multiple independent sources. Our approach assumes that all sources share identical parameter estimates, and we do not currently impose a spatial-dependent structure on the model. Additionally, we adapt the algorithm to handle scenarios with limited data that may not be suitable for asymptotic parametric methods.

\subsection{The Modified Beta-Context Algorithm}



The likelihood of the data $Y_1^n$, conditioning upon knowing $y_1^{\eta}$, $\boldsymbol{x}_1^{\eta}$ and $\boldsymbol{z}$, based on Definition \ref{def_context_modified_vlmcx} under a beta-context tree of order $\eta$ is

\begin{eqnarray}
\label{likelihood_modified_vlmcx}
    \lefteqn{ L(\tau_{\boldsymbol{\theta}} \mid y_1^n, \boldsymbol{x}_1^n, \boldsymbol{z}) = \prod_{s=1}^{q} L(\tau_{\boldsymbol{\theta}} \mid y_1^n (s), \boldsymbol{x}_1^n(s), \boldsymbol{z}(s)) =}  \nonumber \\
    & = &  \prod_{s=1}^{q} P(Y_1^n (s) = y_1^n (s) \mid Y_1^{\eta}(s)  = y_1^{\eta}(s) , \boldsymbol{x}_1^{n} (s), \boldsymbol{z}(s), \tau_{\boldsymbol{\theta}})  \\
    & = & P(Y_{\eta+1} = y_{\eta+1} \mid Y_1^{\eta} = y_1^{\eta}, \boldsymbol{x}_1^{\eta}, \boldsymbol{z}, \tau_{\boldsymbol{\theta}}) \times \ldots \nonumber \\
    & & \quad \times  P(Y_{n} (s) = y_{n} (s) \mid Y_{n-\eta}^{n-1} (s) = y_{n-\eta}^{n-1} (s), \boldsymbol{x}_{n-\eta}^{n-1} (s), \boldsymbol{z} (s), \tau_{\boldsymbol{\theta}}), \nonumber
\end{eqnarray}

\noindent where each probability follows from \eqref{prob_modified_vlmcx}. Notice that this likelihood function is conditioned on knowing the $\eta$ first terms of the chain, this is necessary in the likelihood ratio test in Step 2 to keep the terms in the ratio comparable. For each context $u = y_{-k+1}^0 = y_0, y_{-1}, ..., y_{-k+1}$ with $k$ steps into the past there are $(1+dk+m)$ parameters to estimate for each $j = 1, ..., p-1$. This includes one parameter corresponding to $\alpha_j^u$, $dk$ parameters associated with the $d$ time-varying exogenous covariates at each of the $k$ steps, and $m$ parameters associated with the $m$ time-invariant exogenous covariates. 

Given data ($Y_1^n(s), \boldsymbol{x}_1^n(s), \boldsymbol{z}(s)$) for each independent source $s = 1, ..., q$, the algorithm follows these steps:

\begin{enumerate}
     \item \textbf{Step 1 - Finding Maximal Context Tree:}
     \begin{itemize}
        \item Similar to Step 1 of beta-context algorithm proposed by \citet{zanin2022variable}, initiate the construction of the largest possible beta-context tree, denoted as $\tau_{\max}$. The tree should include all contexts $u \in \tau_{\max}$ that have been observed at least $f(p-1)$ times. The parameter $f$ (where $f \geq 1$) ensures an adequate number of observations per state for estimation. In essence, $\tau_{\max}$ is the most extensive tree meeting the observation criterion. Let

        \begin{align*}
            \tau_{\max} = \{u = y_{-k+1}^0 : N(u) \geq f(p-1)\}
        \end{align*}
    
        where      

        \begin{align*}
            N(u) = \sum_{s=1}^{q}\sum_{t=1}^{n- |u| + 1}I(Y_t^{t+ \abs{u} -1}(s) = u), \ u \in {\cal Y}^{\infty}.
        \end{align*}
        
        Note that, given the presence of multiple sources, $N(u)$  represents the sum of occurrences of the sequence $u$ in each $Y_1^n(s)$, $s = 1, ..., q$.
        
        Define the beta-context function $c(\cdot)$ to form the corresponding context tree $\tau_{\max}$:
        
        \item Set the initial beta-context tree as $\tau^{(0)} = \tau_{\max}$. The variable $r$ denotes the order of this context tree and the initial tree $\tau^{(0)}$ may not be complete.
        
        \item Compute $\hat{\tau}_{\boldsymbol{\theta}}^{(0)}$, the associated estimated parameter tree, using maximum likelihood estimation. Unlike the beta-context algorithm, we allow for the possibility that some parameters may not be estimated. For each context $u$, proceed as follows:
        
        \begin{itemize}
            \item If $N(uj) \geq f(1+d\abs{u}+m) \ \forall j \in {\cal Y}$, calculate all the parameters ($\alpha_j^u$, $d \abs{u}$ parameters associated with the $d$ time-varying exogenous covariates at each of the $\abs{u}$ steps, and $m$ parameters associated with the $m$ time-invariant exogenous covariates). Estimate the parameters by maximizing the likelihood function 

            \begin{align*}
                \prod_{s=1}^q L(\tau_{\boldsymbol{\theta}}^{(0)}\mid y_1^n(s), \boldsymbol{x}_1^n(s), \boldsymbol{z}(s))
            \end{align*}
            
            \item If $N(uj) \geq f(1+m) \ \forall j \in {\cal Y}$ and $N(uj) < f(1+d\abs{u}+m)$ for some $j \in {\cal Y}$, calculate only $\alpha_j^u$ and $m$ parameters associated with the $m$ time-invariant exogenous covariates. Estimate the parameters by maximizing the likelihood function 

            \begin{align*}
                \prod_{s=1}^q L(\tau_{\boldsymbol{\theta}}^{(0)}\mid y_1^n(s), \boldsymbol{z}(s)) = L(\tau_{\boldsymbol{\theta}-\boldsymbol{\beta}} \mid y_1^n, \boldsymbol{x}_1^n, \boldsymbol{z}) 
            \end{align*}
         
            where $\theta - \boldsymbol{\beta}$ is the vector of parameters with all coefficients $\boldsymbol{\beta}$ corresponding to the covariates $\boldsymbol{x}_1^n$ equal to zero. 
            
            \item If $0 < N(uj) < f(1+m)$ for some $j \in {\cal Y}$, calculate only $\alpha_j^u$. Estimate the parameters by maximizing the likelihood function 

            \begin{align*}
                \prod_{s=1}^q L(\tau_{\boldsymbol{\theta}}^{(0)}\mid y_1^n(s)) = L(\tau_{\boldsymbol{\theta}-\boldsymbol{\beta}-\boldsymbol{\gamma}} \mid y_1^n, \boldsymbol{x}_1^n, \boldsymbol{z})
            \end{align*}

            where $\boldsymbol{\theta} - \boldsymbol{\beta} - \boldsymbol{\gamma}$ is the vector of parameters with all coefficients $\boldsymbol{\beta}$ and $\boldsymbol{\gamma}$ corresponding to the covariates $\boldsymbol{x}_1^n$ and $ \boldsymbol{z}$, respectively, equal to zero.
            
            \item If $N(uj) = 0$ for some $j \in {\cal Y}$, calculate only $\alpha_j^u$. Estimate the parameters by maximizing the likelihood function $\prod_{s=1}^q L(\tau_{\boldsymbol{\theta}}^{(0)}\mid y_1^n(s))$ using a single-hidden-layer neural network with a softmax function. In a neural network, we initialize estimates with random values and iteratively adjust them to approximate the real ones. This allows us to estimate parameters even for events that never occur, resulting in estimates for very low probabilities. This differs from traditional Generalized Linear Models (GLMs), which cannot estimate parameters for events that never happen. For a more comprehensive understanding, please refer to \citet{venables2002modern}.
        \end{itemize}
        
        \item It is worth noting that while the initial tree $\tau^{(0)}$ may not be complete, all terminal nodes must possess all possible children. This constraint is imposed because, in the presence of children, we refrain from estimating parameters for the parent node. Consequently, without this constraint, there might be insufficient occurrences for a child node to be represented in the tree, leading to a lack of parameter estimates for that particular node.
    \end{itemize}
    
     \item \textbf{Step 2 - Pruning the Tree (Influence of time-dependent exogenous covariates):}
    \begin{itemize}
        \item For each context $u$ in the initial tree $\tau^{(0)}$ with length $r$, where parameters associated with the $d$ time-varying exogenous covariates have been estimated, apply a likelihood ratio test to examine the significance of the parameter vector associated with past covariates from the node $(-r+1)$ to any of the outcomes (excluding the baseline). This test is performed similarly to Step 2 of the beta-context algorithm  \citep{zanin2022variable},

        \begin{align*}
            H_0^u: \boldsymbol{\beta}_{j, (-r+1)}^u = 0, j = 1, ..., p-1.
        \end{align*}
 
         with the deviance statistic $\lambda_{-r+1}^u$ for testing $H_0^u$ defined as:
         
        \begin{equation}
            \label{test_statistic_betas_modified_vlmcx}
            \begin{aligned}
            \lambda_{-r+1}^u = & -2 \sum_{s=1}^q[\text{log} L( \Tilde{\tau}_{\boldsymbol{\theta}}^u \mid y_1^n(s), \boldsymbol{x}_1^n(s), \boldsymbol{z}(s)) \\ &- \text{log} L(\hat{\tau}_{\boldsymbol{\theta}}^{(0)} \mid y_1^n(s), \boldsymbol{x}_1^n(s), \boldsymbol{z}(s))]
            \end{aligned}
        \end{equation}
        
        where $L(\cdot)$ represents the likelihood function as defined in \eqref{likelihood_modified_vlmcx}.
        
        \item Calculate estimators
        \begin{equation}
        \label{tree_estimator_vlmcx}
           \begin{aligned}
                \Tilde{\tau}_{\boldsymbol{\theta}}^u & = \{(w, \Tilde{\boldsymbol{\theta}}^w): w \in \tau^{(0)}, w \neq u\} \\
                & \cup \{(u, (\Tilde{\alpha}_j^u, \Tilde{\boldsymbol{\beta}}_{j, 0}^u, \Tilde{\boldsymbol{\beta}}_{j, -1}^u, ..., \Tilde{\boldsymbol{\beta}}_{j, (-r+2)}^u, 0),\boldsymbol{\gamma}_j^u), j = 1, ..., p-1\}
            \end{aligned}
        \end{equation}
        
        by maximizing the likelihood under the null hypothesis $H_0^u$.
        \item Compute the p-value $\pi_{-r+1}^u = 1-\Psi_{d(p-1)}(\lambda_{-r+1}^u)$,  where $\Psi_{d(p-1)}(\cdot)$ is the cumulative distribution function of a $\chi^2$ random variable with $d(p-1)$ degrees of freedom.
        
        \item If $\pi_{-r+1}^u > \delta_n$, where $\delta_n$ is a chosen significance level, update the estimated parameter tree $\hat{\tau}_{\boldsymbol{\theta}}^{(0)}$ with $\Tilde{\tau}_{\boldsymbol{\theta}}^{u}$.
    \end{itemize}
    
    \item \textbf{Step 3 - Pruning the Tree (Influence of the context and time-independent exogenous covariates):}
    
    Concerning the tests performed in \textbf{Step 2}:
    \begin{enumerate}
        \item \textbf{Substep 3.1} - If at least two $H_0^{u_k}$ ($k = 1, ..., p$) for $u_1, ..., u_p$ siblings in $\tau_{(0)}$ were not rejected in Step 2, or if they did not have parameters associated with the $d$ time-varying exogenous covariates estimated in Step 1, and any children in this node were not lumped together yet:
        \begin{itemize}
            \item The nodes corresponding to the rejected tests are retained in the model.
            \item Calculate the test statistic for each pair of siblings $(u_i, u_j)$ ($i, j \in \{1, ..., p\}$), where $H_0^{u_i}$ and $H_0^{u_j}$ were not rejected in Step 2, or $u_i$ and $u_j$ did not have parameters associated with the $d$ time-varying exogenous covariates estimated in Step 1:
            \begin{itemize}
                \item If $N(u_ia) \geq f$ and $N(u_ja) \geq f$, $\forall a \in {\cal Y}$ and $f$ defined in Step 1,  
                
                \begin{equation}
                    \label{test_statistic_context_modified_vlmcx}
                    \begin{split}
                    \lambda^{u_i, u_j} = -2\sum_{s=1}^q[\text{log} L(\Tilde{\tau}_{\boldsymbol{\theta}}^{\Tilde{u}}\mid y_1^n(s), \boldsymbol{x}_1^n(s), \boldsymbol{z}(s)) - \\
                    \text{log} L(\hat{\tau}_{\boldsymbol{\theta}}^{(0)} \mid y_1^n(s), \boldsymbol{x}_1^n(s), \boldsymbol{z}(s))],
                    \end{split}
                \end{equation}
                
                \noindent where $\Tilde{\tau}_{\boldsymbol{\theta}}^{\Tilde{u}} = \{(w, \Tilde{\boldsymbol{\theta}}^w): w \in \tau^{(0)}, w \neq u_i, u_j\} \cup \{(\Tilde{u}, \Tilde{\boldsymbol{\theta}}^w):
                \Tilde{u}=(u_i \cup u_j)u, \ u \ \text{is parent of} \ u_1, ..., u_p\}$ is estimated under the null hypothesis that the sibling nodes $u_i$ and $u_j$ are lumped together, reducing the parameters. The parameters in vectors $\boldsymbol{\theta}^{u_i}$ and 
                $\boldsymbol{\theta}^{u_j}$, which have been together reduced to size $\mathbb{R}^{2(1+d(r-1)+m)(p-1)}$ in Step 2, are merged into parameters $\boldsymbol{\theta}^{\Tilde{u}}$ of size $\mathbb{R}^{(1+d(r-1))(p-1)}$.
                
                \item If ($N(u_ia) < f$ and $N(u_ja) \geq f$)  or ($N(u_ja) < f$ and $N(u_ia) \geq f$), for some $a \in {\cal Y}$, and one of the nodes $u_i$ or $u_j$ has estimates for the exogenous covariates parameters, then the corresponding exact test statistic for the likelihood ratio test is the Cochran-Mantel-Haenszel statistic (for additional details refer to \cite{agresti2012categorical}). To compute this statistic, for the node that lacks covariate estimates, sum the event occurrences across all sources. For the other node, segregate event occurrences for each source. Subsequently, each contingency table will represent a source, and the row of the node without exogenous covariate estimates will be consistent across all contingency tables.
                \item If $N(u_ia) < f$ and $N(u_ja) < f$, for some $a \in {\cal Y}$, the corresponding exact test statistic for the likelihood ratio test is the Fisher statistic (for additional details refer to \cite{agresti2012categorical}). To compute the Fisher statistic, aggregate the event occurrences across all sources for both nodes, resulting in a single contingency table. Each row in this table corresponds to a node.
            \end{itemize}

            \item For the smallest $\lambda^{u_i, u_j}$, compute the p-value $\pi^{(u_i, u_j)}$. If $N(u_ia) \geq f$ and $N(u_ja) \geq f$, $\forall a \in {\cal Y}$, $\pi^{(u_i, u_j)}  = 1 - \Psi_{(1+d(r-1))(p-1)}(\lambda^{(u_i, u_j)})$. If $\pi^{(u_i, u_j)} \geq \delta_n$, lump siblings nodes $(u_i, u_j)$ together in $\Tilde{u}$. Update both $\tau^{(0)}$ and $\hat{\tau}_{\boldsymbol{\theta}}^{(0)}$ with $\Tilde{\tau}_{\boldsymbol{\theta}}^{\Tilde{u}}$. 
        \end{itemize}
        
        \item \textbf{Substep 3.2} - If any two children were lumped together in Substep 3.1 and still have at least one of $H_0^{u_k} (k = 1, ..., p)$ not rejected in Step 2, where $u_k$ were not lumped together in $\Tilde{u}$ and $u_1, ..., u_p$ siblings:
        
        \begin{itemize}
            \item The nodes corresponding to the rejected tests are retained in the model.
            
            \item Calculate the test statistic $\lambda^{u_k, \Tilde{u}}$ (following the guidelines provided in Substep 3.1) for each $u_k \ (k = 1, ..., p)$ which had not been lumped together in $\Tilde{u}$ and had $H_0^{u_k}$ not rejected in Step 2.
            
            \item For the smallest $\lambda^{u_k, \Tilde{u}}$ compute the p-value $\pi^{(u_k, \Tilde{u})}$ (following the guidelines provided in Substep 3.1). If $\pi^{(u_k, \Tilde{u})} \geq \delta_n$, lump node $u_k$ together in $\Tilde{u}$. Update both $\tau^{(0)}$ and $\hat{\tau}_{\boldsymbol{\theta}}^{(0)}$. 
            
            \item Repeat Substep 3.2 until no more $u_k \ (k = 1, ..., p)$ can be lumped together in $\Tilde{u}$.
        \end{itemize}
        
        \item \textbf{Substep 3.3} Lastly, sequentially test to prune the past most parameters in $\boldsymbol{\beta}^{u_k}$ ($k = 1, ..., p$) which had its hypothesis not rejected in Step 2 up to the root, similar to Substep 3.2 of the beta-context algorithm. Calculate the corresponding test statistic $\lambda_{-r+2}^{u_1}$ using \eqref{test_statistic_betas_modified_vlmcx}.
        
    \end{enumerate}
    
    At the end of Step 3, the tree structure and parameters are possibly pruned based on the significance of context and covariate influence, leading to a potentially smaller context tree $\tau^{(1)} \subseteq \tau^{(0)}$ and its updated parameter tree $\hat{\tau}_{\boldsymbol{\theta}}^{(1)}$ at level $r$.
    
    \item \textbf{Step 4 - Iterative Pruning:}
    \begin{itemize}
        \item Repeat Steps 2 and 3 for contexts of lengths $r-1, r-2, ..., 1$, using the updated trees $\tau^{(1)}$ and $\hat{\tau}_{\boldsymbol{\theta}}^{(1)}$.
        
        \item However, if $u_1, ..., u_p$ are siblings and at least one of them has children, no context pruning is performed (both nodes are retained). But, as in Step 3, the pruning of covariate parameters is executed sequentially, proceeding from the most distant past to the root.
    \end{itemize}
\end{enumerate}

Denote this pruned beta-context tree by $\hat{\tau}_n$ with associated parameter tree $\hat{\tau}_{\boldsymbol{\theta}}$ and corresponding beta-context function $\hat{c}(\cdot)$.

It can be shown, following a methodology similar to \citet{zanin2022variable}, that the modified beta-context algorithm ensures strong consistency in estimating the beta-context tree, regression parameters, and transition probabilities.

\section{Predicting Dengue outbreaks}

We applied the proposed modified VLMCX model with time-varying and time-invariant exogenous covariates for dengue outbreak prediction in two distinct scenarios. The first one studies monthly dengue cases from 126 municipalities in Brazil from January 2008 to July 2023 including time-dependent climate factors and time-invariant socioeconomic and demographic attributes as covariates to investigate the influence of previous dengue rates, weather conditions, and socioeconomic factors on subsequent dengue rates across diverse municipalities, providing insights into dengue transmission dynamics. The current model does not incorporate a spatial-dependent structure. However, considering that mosquitoes can fly over a range of 100-500 meters \citep{siriyasatien2018dengue}, municipalities are assumed to be independent. 

As pointed out before, machine learning algorithms were used by  \citet{aleixo2022predicting} to model dengue spread. Thus, to compare our model to theirs, our second scenario studies a historical data series of monthly dengue cases for each district in the city of Rio de Janeiro, Brazil in the period January 2012 to September 2020.  The dataset utilized in this analysis is the same as the one used by \citet{aleixo2022predicting} \footnote{The data used are available at \url{https://gitlab.com/interscity/health/dengue-prediction}. Accessed August 13, 2024.}. In this case, the covariates included in the model were time-dependent variables like temperature, precipitation, air humidity, the number of Chikungunya and Zika cases, and the \textit{Aedes aegypti} infestation index. Time-invariant covariates include fixed attributes of districts, such as the sum of dengue cases in neighboring districts, demographic density, and the number of health facilities. Although the current model does not incorporate a spatial-dependent structure,  the sum of dengue cases in neighboring districts in the past months is incorporated as an exogenous covariate in the model.

\subsection{Scenario 1: National Analysis - Brazil (January 2008 to July 2023)}

To explore the analysis of dengue transmission across multiple municipalities in Brazil, we collected information on monthly dengue notifications recorded in the Notifiable Diseases Information System (Sistema de Informa\c c\~ao de Agravos de Notifica\c c\~ao - SINAN) \footnote{The data used are available at \url{https://datasus.saude.gov.br/informacoes-de-saude-tabnet/}. Accessed August 13, 2024} from January 2008 to July 2023. From the literature review of main factors correlated with the number of dengue cases we chose to study {\bf climate factors} (monthly average temperature (\degree{C}), monthly rainfall (mm), and monthly total days of rainfall), {\bf socioeconomic factors} (percentage of the population in a low-income situation ($< 1/2$ minimum wage),  Gross Domestic Product (GDP) per Capita, and percentage of the population with access to a sewage system) and {\bf demographic factors} (percentage of the population living in urban areas according to the national demographic census from 2010, area of the municipality ($\text{km}^2$), and total population of the municipality according to the national demographic census from 2010 and 2022. For climate data, we use records from automatic weather stations spanning January 2008 to July 2023, provided by the National Institute of Meteorology (Instituto Nacional de Meteorologia - INMET).\footnote{\url{https://portal.inmet.gov.br/dadoshistoricos}. Accessed August 13, 2024} Socioeconomic and demographic data are obtained from the official statistics bureau, the Brazilian Institute of Geography and Statistics (Instituto Brasileiro de Geografia e Estatística - IBGE).\footnote{\url{https://www.ibge.gov.br/geociencias/organizacao-do-territorio/estrutura-territorial/15761-areas-dos-municipios.html}. Accessed August 13, 2024}\footnote{\url{https://sidra.ibge.gov.br/pesquisa/censo-demografico/demografico-2010/inicial}. Accessed August 13, 2024}\footnote{\url{https://sidra.ibge.gov.br/pesquisa/censo-demografico/demografico-2022/inicial}. Accessed August 13, 2024}
  
First, we considered only municipalities with complete information on dengue incidence from 2008 to 2023, totaling 1179 municipalities. Out of these, only 280 had identified automatic weather stations and were considered in the sample. For these 280 automatic weather stations, we applied a moving average approach for each year and station to address missing values in temperature and rainfall data. This was done exclusively for automatic weather stations with at least seven months of complete information in the corresponding year. For any municipality with years lacking more than four months of data, we excluded those years and used weather data from the years either preceding the first incomplete year or following the last incomplete year, depending on which option resulted in a more comprehensive dataset. At the end of this approach, we had 237 municipalities. Finally, we included only municipalities with at least 36 months of weather information, totaling 126 municipalities from all regions of Brazil.

Based on a descriptive analysis and literature review, we decided to include the following exogenous covariates in the model fitting:

\begin{itemize}
  \item \textbf{Time-varying exogenous covariates:} Average monthly temperature and monthly days of precipitation.
  \item \textbf{Time-invariant exogenous covariates:} Percentage of population in a low-income situation (2010) divided by 10 and percentage of the population living in urban areas (2010). We divided these covariates by 10 to standardize their scale and ensure that the exogenous variables are more comparable. 
\end{itemize}

\begin{figure}[h!]
\begin{center}
\includegraphics[scale=0.3]{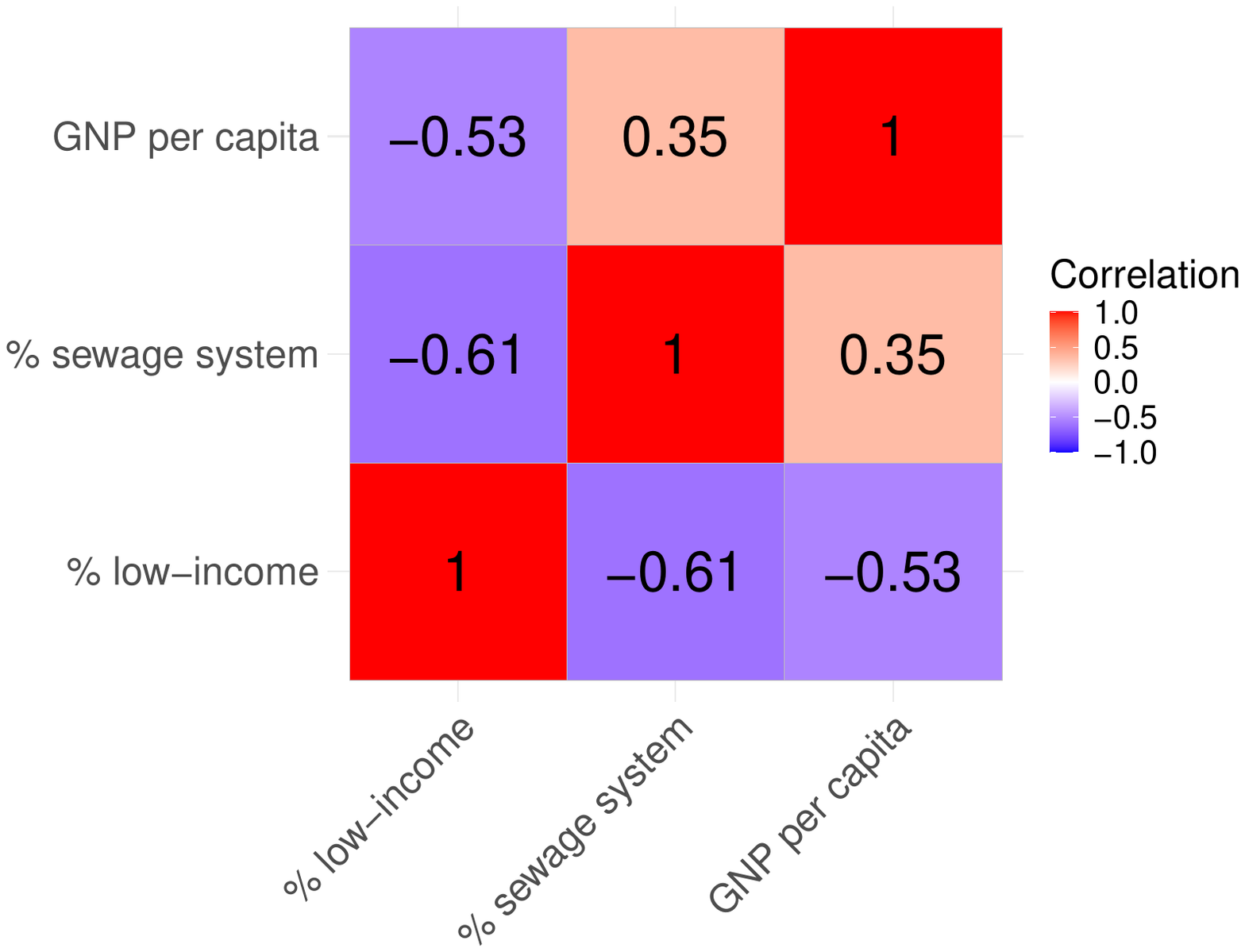}
\caption{\label{correlation} Correlation matrix between socioeconomic factors}
\end{center}
\end{figure}

The choice of using days of precipitation instead of total precipitation stems from the fact that mosquitoes require stagnant water to complete their life cycle \citep{buczak2012data}. Therefore, it is not only about the volume of rain but also the constancy of rainfall over time, allowing for the presence of standing water after warm periods. Additionally, the decision to use the percentage of the population in a low-income situation instead of the percentage of the population with a sewage system or GDP per capita is based on the belief that it provides a better representation of municipal poverty and social inequalities. GDP per capita can be influenced by individuals with incomes well above the average while having or lacking a sewage system is not always indicative of high or low-income situations in contemporary settings. Moreover, the percentage of the population in a low-income situation shows a stronger correlation with the other two social factors (\autoref{correlation}).

The monthly dengue cases were categorized based on their distribution (\autoref{dist_dengue}) and the Ministry of Health's classification of dengue incidence (cases per 100,000 inhabitants): 

\begin{itemize}
  \item \textbf{Category 1} - Up to 5 cases per 100,000 inhabitants, 
  \item \textbf{Category 2} - 5 to 25 cases per 100,000 inhabitants, 
  \item \textbf{Category 3} - 25 to 75 cases per 100,000 inhabitants, and 
  \item \textbf{Category 4} - Over 75 cases per 100,000 inhabitants.
\end{itemize}

\begin{table}[h!]
\centering
\caption{Distribution of dengue cases for all the 126 municipalities. Q1 and Q3 represent the first and third quartiles respectively.}
\label{dist_dengue}
\begin{tabular}{l|cccccc}
\multicolumn{1}{c|}{Situation} & Min & Q1 & Median & Mean  & Q3 & Max \\ \hline
Epidemic                       & 0       & 3.86                     & 17.57  & 98.31 & 71.84                    & 4478.30 \\
Non-epidemic                   & 0       & 0                        & 0.89   & 2.93  & 3.53                     & 56.76  
\end{tabular}%
\end{table}

To prepare the data for modeling, we excluded the most recent 12 months of data from each municipality, reserving them for prediction. For this fitting, we chose to employ multinomial logistic regression for parameter estimation in each context. Although we also experimented with ordinal logistic regression and found similar forecasting results, we opted to present the results for multinomial logistic regression fitting, as we believe it provides clearer estimations.

For tuning parameters $\delta$ and $f$, the significance level and the minimum number of observations per state for estimation, respectively, we selected values that minimized the Bayesian Information Criterion (BIC), resulting in $\delta = 0.000001$ and $f = 4$. Additionally, we constrained the initial tree to have a maximum depth of 6, with Category 1 assumed as the baseline.

\begin{figure}[h!]
    \begin{center}
        \noindent
        $\begin{aligned}
          \begin{forest}
              [y 
                      [1
                          [1  
                              [1]]]
                      [2
                          [1]]
                      [3 
                          [1]
                          [2]]
                      [4
                          [1]]]
          \end{forest}
          \end{aligned}$
        \noindent
    \end{center}
    \caption{\label{final_tree_br} Final tree for dengue incidence in Brazilian municipalities}
\end{figure}
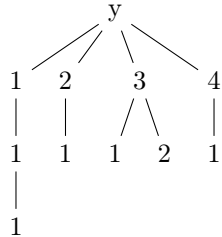

The estimated beta-context is depicted in \autoref{final_tree_br} and the corresponding parameter estimates are in \autoref{final_tree_br_estimates}. For the parameters $\hat{\alpha}$ and $\hat{\boldsymbol{\gamma}}$, the subscript represents the next state, while the superscript denotes the context. For example, $\hat{\alpha}_{2}^{111}$ and $\hat{\boldsymbol{\gamma}}_{2}^{111}$ represent the estimates of the intercept and time-invariant covariates for the transition probability from context 111 to state 2.

For $\hat{\boldsymbol{\beta}}$, the first subscript corresponds to the next state, the second subscript indicates the time point in the past considered for the covariates' influence, and the superscript represents the context. For example, $\hat{\boldsymbol{\beta}}_{2,1}^{111}$ represents the estimate for the most recent past of the time-dependent covariates affecting the transition probability from context 111 to state 2. Although the context considers three past time points, covariate parameter estimates are available only for the most recent one.

Additionally, the first element of $\hat{\boldsymbol{\beta}}$ corresponds to estimates for the average monthly temperature, while the second represents estimates for the monthly number of days with precipitation. For $\hat{\boldsymbol{\gamma}}$, the first element represents estimates for the percentage of the population in a low-income situation, and the second represents estimates for the percentage of the population living in urban areas.

Contexts 1* \footnote{1* means the context is 1 preceded by other states that are not 1}, 11, and 41 lacked sufficient observations to estimate exogenous covariate parameters. Across all contexts, only time-varying exogenous covariates of the last month demonstrated a significant effect, with estimates for previous months being equal to zero.


\begin{table}[h!]
    \caption{Estimated parameters for dengue incidence in Brazilian municipalities}
    \label{final_tree_br_estimates}
    \centering
    \resizebox{\textwidth}{!}{%
    \begin{tabular}{|c|c|c|c|}
        \hline
        \textbf{\begin{tabular}[c]{@{}c@{}}Contexts/\\ Estimators\end{tabular}} &
        \textbf{$\hat{\alpha}$} &
        \textbf{$\hat{\boldsymbol{\beta}}$} &
        \textbf{$\hat{\boldsymbol{\gamma}}$} \\ \hline
        \multirow{3}{*}{\textbf{1*}} &
        $\hat{\alpha}^{1*}_{2} = -1.27$ &
        \multirow{3}{*}{} &
        \multirow{3}{*}{} \\
        &
        $\hat{\alpha}^{1*}_{3} = -2.8$ &
        &
        \\
        &
        $\hat{\alpha}^{1*}_{4} = 4.45$ &
        &
        \\ \hline
        \multirow{3}{*}{\textbf{111}} &
        $\hat{\alpha}^{111}_2 = -6.51$ &
        $\hat{\boldsymbol{\beta}}^{111}_{2,1} = (0.2,0.06)'$ &
        $\hat{\boldsymbol{\gamma}}^{111}_{2} = (-0.01,-0.01)'$ \\
        &
        $\hat{\alpha}^{111}_3 = -7.27$ &
        $\hat{\boldsymbol{\beta}}^{111}_{3,1} = (0.28,0.07)'$ &
        $\hat{\boldsymbol{\gamma}}^{111}_{3} = (-0.03,-0.04)'$ \\
        &  
        $\hat{\alpha}^{111}_4 = -11.88$ &
        $\hat{\boldsymbol{\beta}}^{111}_{4,1} = (0.37,0.06)'$ &
        $\hat{\boldsymbol{\gamma}}^{111}_{4} = (-0.02,-0.03)'$ \\ \hline
        \multirow{3}{*}{\textbf{2*}} &
        $\hat{\alpha}^{2*}_2 = -6.68$ &
        $\hat{\boldsymbol{\beta}}^{2*}_{2,1} = (0.21,0.04)'$ &
        $\hat{\boldsymbol{\gamma}}^{2*}_{2} = (-0.01,0.03)'$ \\
        &
        $\hat{\alpha}^{2*}_3= -6.76$ &
        $\hat{\boldsymbol{\beta}}^{2*}_{3,1} = (0.35,0.09)'$ &
        $\hat{\boldsymbol{\gamma}}^{2*}_{3} = (-0.05,-0.02)'$ \\
        &
        $\hat{\alpha}^{2*}_4 = 8.43$ &
        $\hat{\boldsymbol{\beta}}^{2*}_{4,1} = (0.4,0.1)'$ &
        $\hat{\boldsymbol{\gamma}}^{2*}_{4} = ( -0.06,-0.03)'$ \\ \hline
        \multirow{3}{*}{\textbf{21}} &
        $\hat{\alpha}^{21}_2= -8.1$ &
        $\hat{\boldsymbol{\beta}}^{21}_{2,1} = (0.2,0.07)'$ &
        $\hat{\boldsymbol{\gamma}}^{21}_{2} = (0,0.03)'$ \\
        &
        $\hat{\alpha}^{21}_3 = -8.65$ &
        $\hat{\boldsymbol{\beta}}^{21}_{3,1} = (0.2,0.1)'$ &
        $\hat{\boldsymbol{\gamma}}^{21}_{3} = (-0.01,0.02)'$ \\
        &
        $\hat{\alpha}^{21}_4 = -9.9$ &
        $\hat{\boldsymbol{\beta}}^{21}_{4,1} = (0.28,0.1)'$ &
        $\hat{\boldsymbol{\gamma}}^{21}_{4} = (-0.01,0.01)'$ \\ \hline
        \multirow{3}{*}{\textbf{3*}} &
        $\hat{\alpha}^{3*}_2 = -6.65$ &
        $\hat{\boldsymbol{\beta}}^{3*}_{2,1} = (0.11,0.03)'$ &
        $\hat{\boldsymbol{\gamma}}^{3*}_{2} = (0.01,0.06)'$ \\
        &
        $\hat{\alpha}^{3*}_3 = 11.72$ &
        $\hat{\boldsymbol{\beta}}^{3*}_{3,1} = (0.36,0.08)'$ &
        $\hat{\boldsymbol{\gamma}}^{3*}_{3} = (-0.02,0.05)'$ \\
        &
        $\hat{\alpha}^{3*}_4 = -13.43$ &
        $\hat{\boldsymbol{\beta}}^{3*}_{4,1} = (0.5,0.13)'$ &
        $\hat{\boldsymbol{\gamma}}^{3*}_{4} = (-0.04,0.03)'$ \\ \hline
        \multirow{3}{*}{\textbf{31}} &
        $\hat{\alpha}^{31}_2 = 1.64$ &
        \multirow{3}{*}{} &
        $\hat{\boldsymbol{\gamma}}^{31}_{2} = (-0.02,-0.01)'$ \\
        &
        $\hat{\alpha}^{31}_3 = 8.39$ &
        &
        $\hat{\boldsymbol{\gamma}}^{31}_{3} = (0.05,0.09)'$ \\
        &      
        $\hat{\alpha}^{31}_4 = 3.57$ &
        &
        $\hat{\boldsymbol{\gamma}}^{31}_{4} = (-0.06,-0.01)'$ \\ \hline
        \multirow{3}{*}{\textbf{32}} &
        $\hat{\alpha}^{32}_2 = 2.21$ &
        \multirow{3}{*}{} &
        $\hat{\boldsymbol{\gamma}}^{32}_{2} = (-0.03,0)'$ \\
        &
        $\hat{\alpha}^{32}_3 = -1.18$ &
        &
        $\hat{\boldsymbol{\gamma}}^{32}_{3} = (-0.02,0.05)'$ \\
        &
        $\hat{\alpha}^{32}_4 = 1.87$ &
        &
        $\hat{\boldsymbol{\gamma}}^{32}_{4} = (-0.04,0.02)'$ \\ \hline
        \multirow{3}{*}{\textbf{4*}} &
        $\hat{\alpha}^{4*}_2 = -3.19$ &
        $\hat{\boldsymbol{\beta}}^{4*}_{2,1} = (-0.10,-0.04)'$ &
        $\hat{\boldsymbol{\gamma}}^{4*}_{2} = (0.04,0.08)'$ \\
        &
        $\hat{\alpha}^{4*}_3 = -5.84$ &
        $\hat{\boldsymbol{\beta}}^{4*}_{3,1} = (0.07,-0.04)'$ &
        $\hat{\boldsymbol{\gamma}}^{4*}_{3} = (0.03,0.08)'$ \\
        &
        $\hat{\alpha}^{4*}_4 = 9.68$ &
        $\hat{\boldsymbol{\beta}}^{4*}_{4,1} = (0.31,0.04)'$ &
        $\hat{\boldsymbol{\gamma}}^{4*}_{4} = (0,0.08)'$ \\ \hline
        \multirow{3}{*}{\textbf{41}} &
        $\hat{\alpha}^{41}_2 = 0$ &
        \multirow{3}{*}{} &
        \multirow{3}{*}{} \\
        &
        $\hat{\alpha}^{41}_3 = 0.09$ &
        &
        \\
        &
        $\hat{\alpha}^{41}_4 = 2.6$ &
        &
        \\ \hline 
        \multicolumn{4}{c}{%
            \begin{minipage}{17cm}%
                \footnotesize
                \vspace{1mm}\textbf{Note:} $\hat{\alpha}_{j}^{u}$ represents the estimated intercept,  $\hat{\boldsymbol{\gamma}}_{j}^{u}$ the estimated coefficients for the time-invariant covariates (with coordinates corresponding to the percentage of the population in low-income situations and the percentage living in urban areas, respectively) and, $\hat{\boldsymbol{\beta}}_{j,t}^{u}$ represents the estimated coefficients for time-dependent covariates (with coordinates corresponding to the average monthly temperature and the monthly number of days with precipitation, respectively) that influence the transition probability from context $u$ to state $j$ at time $t$.
            \end{minipage}%
        }
    \end{tabular}%
    }
\end{table}

The estimated context tree given by Figure \ref{final_tree_br} provides valuable insights into the trend of dengue incidence. According to the tree, when there are less than 5 cases per 100,000 inhabitants in the previous month, the next month's dengue incidence appears to be independent of the rest of the history, except when there are also fewer than 5 cases per 100,000 inhabitants in the history. In other words, when examining the preceding month, if there are fewer than 5 cases per 100,000 inhabitants, it is necessary to consider an additional previous month. If there are still fewer than 5 cases per 100,000 inhabitants, it becomes necessary to look back three steps in the past. Conversely, if there are more than 5 cases per 100,000 inhabitants, there is no need to look further back than one month ago. This observation suggests that having less than 5 cases per 100,000 inhabitants generally indicates a decreasing trend, and the specific events preceding that period may not significantly impact the subsequent month's incidence, except when there is a continued low incidence. Similarly, when there are 25 to 75 cases per 100,000 inhabitants in the previous months, dengue incidence is independent of the remaining history unless there were less than 25 cases per 100,000 inhabitants in the month before last. This pattern could be explained by the likelihood that, when there are already 25 to 75 cases per 100,000 inhabitants in the previous months, the probability is higher that cases are decreasing. However, if there were fewer than 25 cases per 100,000 inhabitants, there might be a chance of the cases increasing.

Regarding time-varying exogenous covariates, it can be observed in Table \ref{final_tree_br_estimates} that, in general, increases in temperature and rainfall correspond to a higher probability of experiencing higher dengue cases. However, concerning time-invariant exogenous covariates, the estimates are notably low, suggesting that there may be no significant impact on dengue incidence. This observation aligns with the discussion by \citet{horstick2015reviewing}, who argue that although dengue predominantly affects resource-limited countries, it does not exclusively target the poor.

\begin{figure}[h!] 
\begin{center}
\includegraphics[scale=0.6]{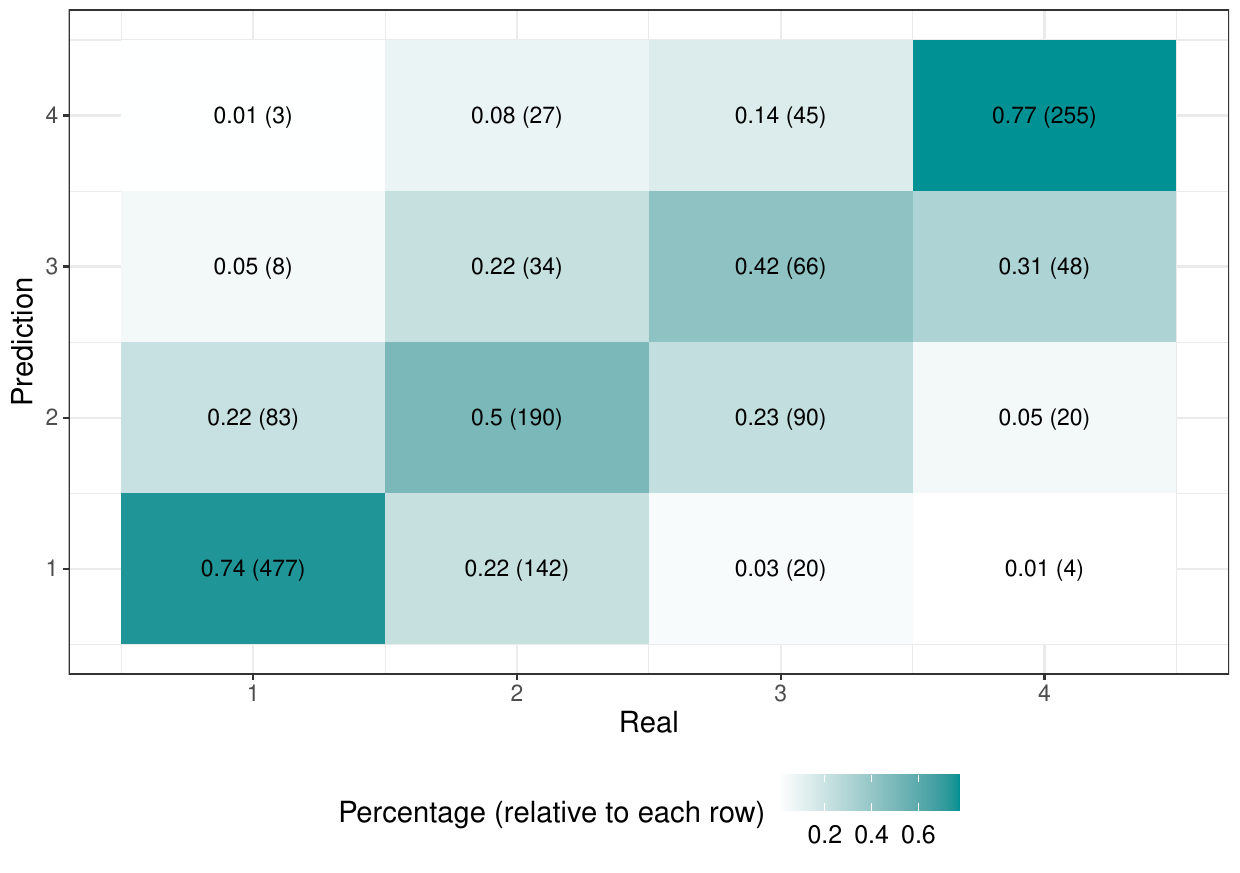}
\end{center}
\caption{\label{confusion_matrix_br}Confusion matrix for prediction results (using all available observations from previous months) - Brazil municipalities}
\end{figure}

\begin{figure}[h!]
\hspace{-1.5cm}\includegraphics[scale=0.45]{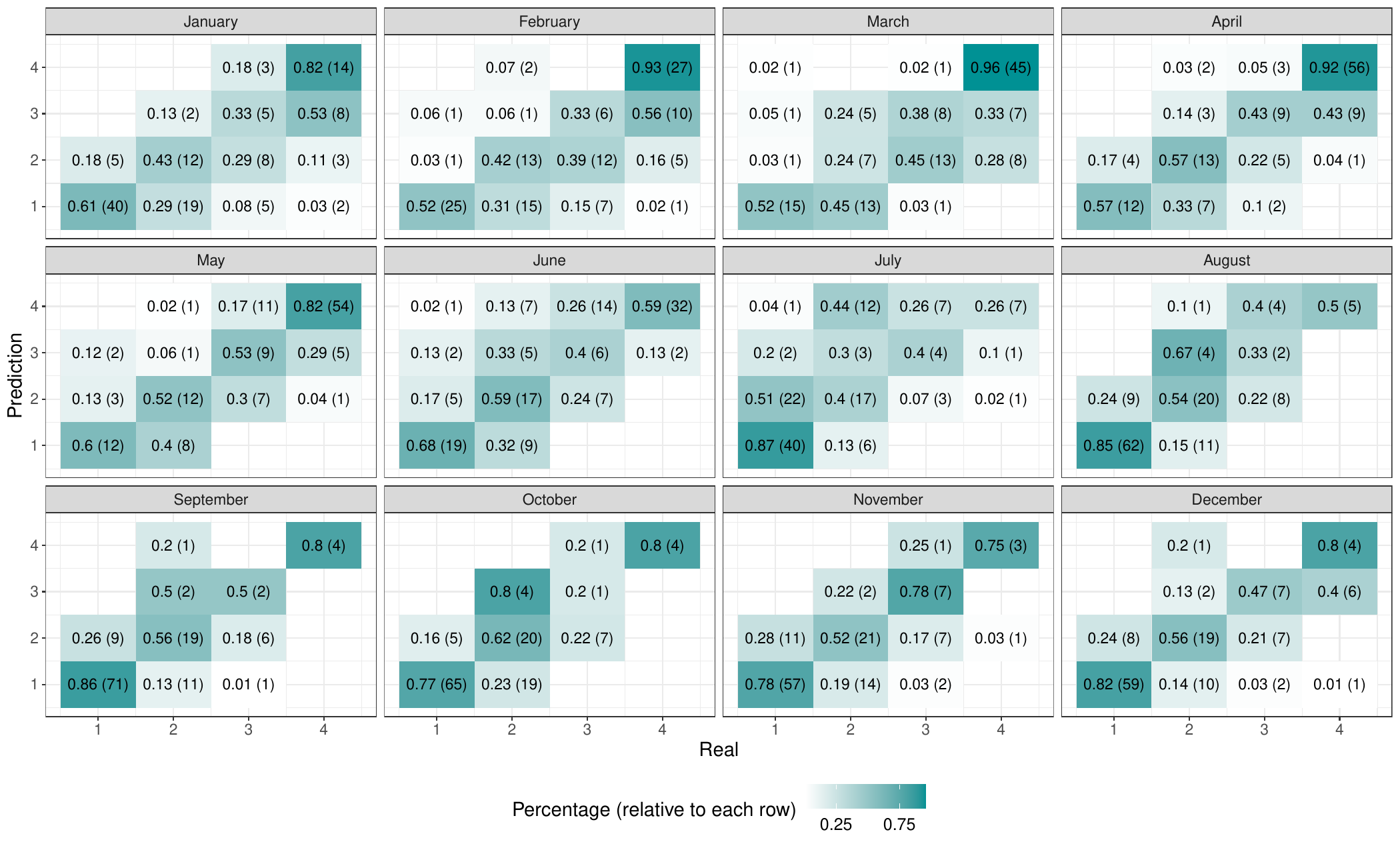}
\caption{\label{confusion_matrix_month_br}Confusion matrix for prediction results by month (using all available observations from previous months) - Brazil municipalities}
\end{figure}

All available observations from previous months were used to
forecast each of the 12 months, as the model relies on covariates from previous months. \autoref{confusion_matrix_br} and \autoref{confusion_matrix_month_br} displays the confusion matrices of forecast results, both in total and broken down by the month of the year.

The model demonstrated satisfactory performance in predicting dengue incidence above 75 monthly cases/100,000 inhabitants, accurately identifying high incidence in 77\% of the cases. It also showed the ability to recognize situations where the actual incidence may be lower than predicted, with 69\% of cases falling within or below the predicted range of 25 to 75 cases/100,000 inhabitants. However, caution is advised when dengue incidence may be higher than initially predicted (31\% for Category 3, 28\% for Category 2, and 26\% for Category 1). 

\begin{figure}[h!]
\begin{center}
\includegraphics[scale=0.6]{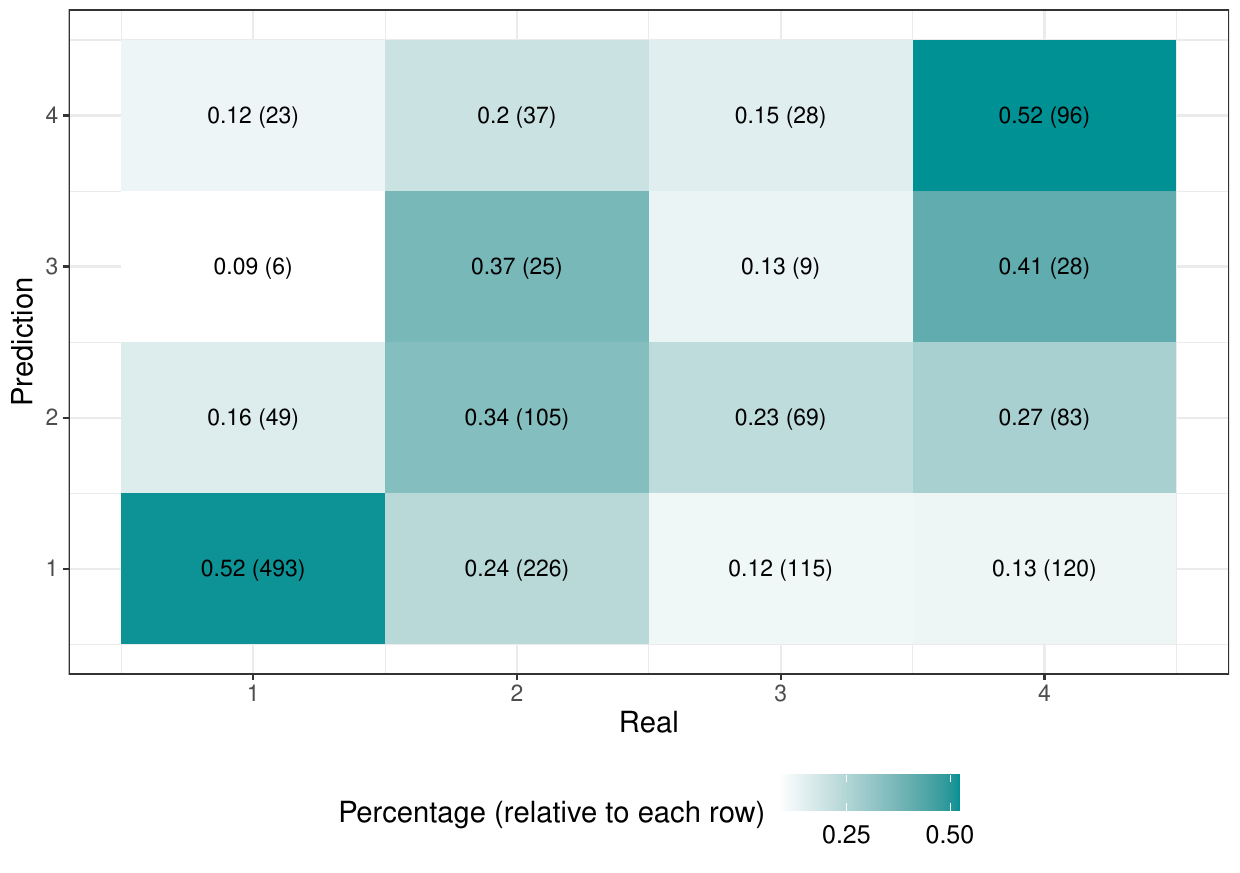}
\end{center}
\caption{\label{confusion_matrix_2_br} Confusion matrix for prediction results (using previously predicted dengue incidence instead of real values) - Brazil municipalities}
\end{figure}

\begin{figure}[h!]
\hspace{-1.5cm}\includegraphics[scale=0.45]{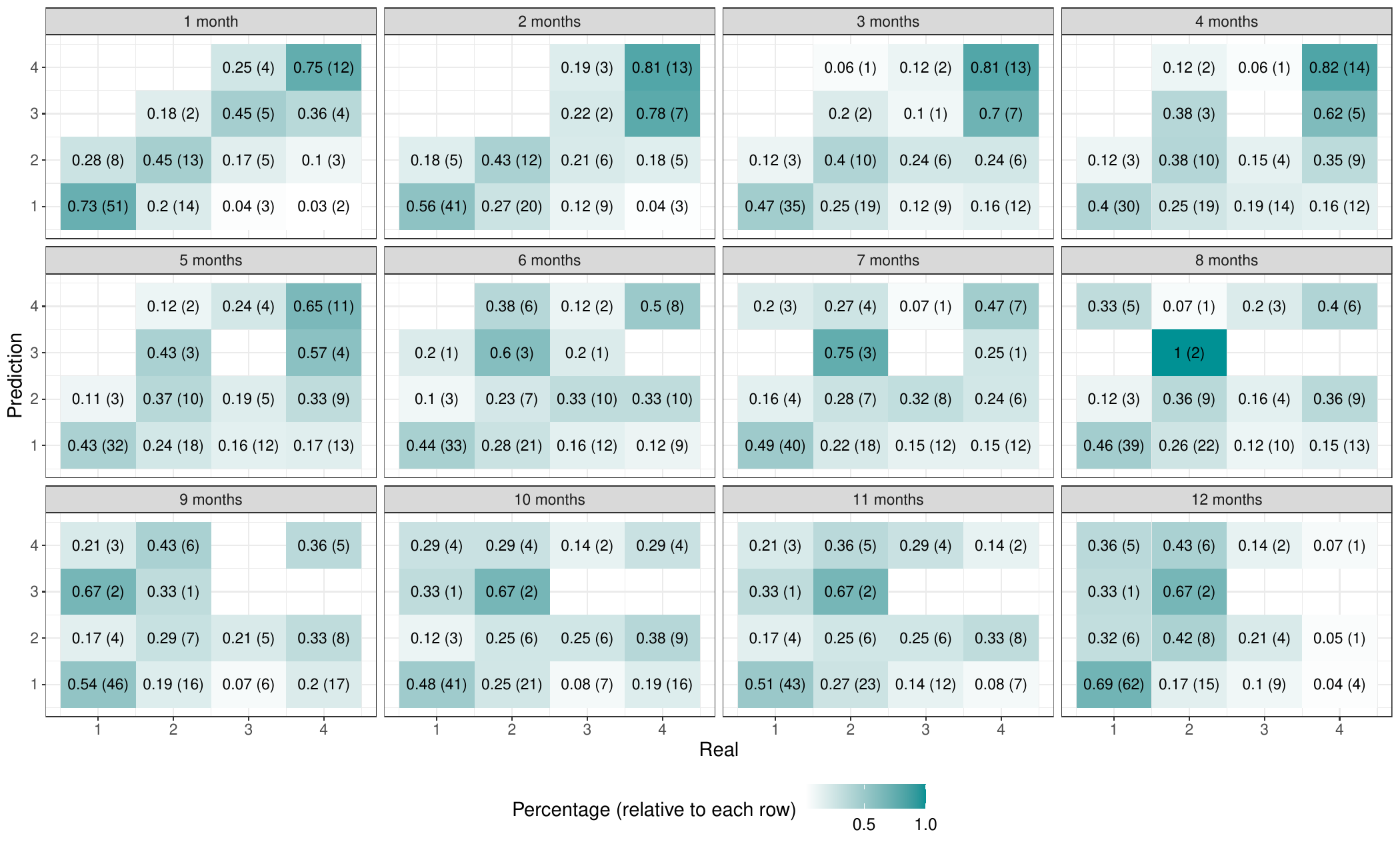}
\caption{\label{confusion_matrix_month_2_br}Confusion matrix for prediction results by time ahead (using previously predicted dengue incidence instead of real values) - Brazil municipalities}
\end{figure}

Recognizing that dengue incidence data may not always be available for the previous month due to the time it takes for consolidation and availability, we also explored predicting using previously predicted dengue incidence instead of real values. In this approach, for the second month of prediction, we utilized the predicted values from the first month and continued this process consecutively. Therefore, to predict 12 months, we used dengue incidence predictions for the previous 11 months instead of the actual dengue incidence in those months. Notice that we compared predictions using  $k$, $k = 1, \ldots, 11$, months of predicted values, and true values for further past.  \autoref{confusion_matrix_2_br} and \autoref{confusion_matrix_month_2_br} present the confusion matrices of forecast results, both in total and broken down by the time ahead forecasted. As municipalities may have a different number of months, the time ahead may not always correspond to the same month of the year in each municipality. It is important to note that for time-varying exogenous covariates, we utilized only real values.

The model exhibited satisfactory performance in predicting dengue incidence up to 2 months ahead but started to deviate more for predictions beyond 2 months, especially for intermediary categories. However, the model demonstrated satisfactory performance in predicting dengue incidence above 75 monthly cases per 100,000 inhabitants, even for more than 4 months ahead.

\subsection{Scenario 2: City-Level Analysis - Rio de Janeiro, Brazil (January 2011 to September 2020)}


For this study, we utilized the dataset provided by \citet{aleixo2022predicting} \footnote{Publicly accessible on GitLab (\url{https://gitlab.com/interscity/health/dengue-prediction}). Accessed August 16, 2024.}. Since the purpose of this scenario is to compare our results with theirs, variable selection was not required, as the same set of variables was used. Additionally, since there was no need to specify how far back to look, we included the number of dengue, Zika, and Chikungunya cases, \textit{Aedes aegypti} infestation index, total precipitation (mm), mean temperature (\degree{C}) and mean air humidity (\%) for all previous months. This approach allowed the model to automatically determine the relevant time lag for each variable. \autoref{tab_features_rj} shows the full list of features.

\begin{table}[h!]
\centering
\caption{Features used as input to the model - Rio de Janeiro, Brazil (January 2012 to October 2020).  Source: \citet{aleixo2022predicting}}
\label{tab_features_rj}
\resizebox{\textwidth}{!}{%
\begin{tabular}{ll}
\hline
\textbf{Feature}             & \textbf{Description}                                                 \\ \hline
cases-n            & Past number of dengue cases, where n = 1, 2, 3 is the number of months in the past (per district) \\
dengue\_prevalence & Sum of dengue cases in the past, normalized to the 0 to 1 range (per district)                    \\
neighbor\_cases     & Sum of dengue cases in neighboring districts (per district) \\
precipitation (mm)  & Total precipitation in the last month (per district)        \\
temperature (\degree{C})    & Mean temperature in the last month (per district)           \\
air\_humidity (\%)  & Mean air humidity in the last month (per district)          \\
liraa               & \textit{Aedes aegypti} infestation index (for the city)              \\
chikungunya          & Number of chikungunya cases last month (per district)       \\
zika                & Number of zika cases last month (per district)              \\
demographic density & Demographic density (per district)                          \\
num\_health\_unit   & Number of health facilities (per district)                  \\ \hline
\end{tabular}%
}
\end{table}

So, in summary, for each district, the modeling setup is as follows:

\begin{itemize}
  \item \textbf{Response variable:} 
  The number of dengue cases, categorized following \citet{aleixo2022predicting}: 
  
  \begin{itemize}
    \item \textbf{Severe outbreak:} when the number of cases in a district in a given month is above 99\% of all training set measurements.
    \item \textbf{Mild outbreak:} when the number of cases in a district in a given month is above 95\% of all training set measurements.
    \item \textbf{No outbreak :} when the number of cases in a district in a given month is below or equal to 95\% of all training set measurements.
  \end{itemize}
   
   \item \textbf{Time-varying exogenous covariates:} Total precipitation (mm), mean temperature (ºC), mean air humidity (\%), the sum of dengue cases in neighboring districts,  number of Zika cases, number of Chikungunya cases, and \textit{Aedes aegypti} infestation index.
   
   \item \textbf{Time-invariant exogenous covariates:} Demographic density, number of health facilities, and the sum of dengue cases in the past (normalized to the 0 to 1 range). 
\end{itemize}

In their study, \citet{aleixo2022predicting} employed a boosted tree regression method (CatBoost) and compared its results with a Seasonal Autoregressive Integrated Moving Average (SARIMA) model, using individual time series for each district. For CatBoost, they conducted a grid search to define tuning parameters, using 2015 as validation data and 2012 to 2014 as training data. In the case of SARIMA, tuning parameters were defined fitting the model to data from 2012 to 2015.

Regarding predictions, individual predictions were made for each district and month in the period 2016 to 2020. For CatBoost, they established a distinct model for each year, employing a 5-fold cross-validation approach. This involved using a single year as the test set and four years as the training set. For instance, when predicting the number of cases in 2017, the training data consisted of the years 2016, 2018, 2019, and 2020. In the case of the SARIMA model, they also created separate models for each year (2016 to 2020), using the previous four years as training data, as SARIMA requires a contiguous time series to extract trend and seasonality features.

To ensure comparability with \citet{aleixo2022predicting}, we also created a separate model for each year (2016 to 2020). However, due to the requirement of a contiguous time series for Markov Chains, we used the previous four years as training data. For example, to predict the number of cases in 2016, we utilized data from 2015, 2014, 2013, and 2012 as training data. Additionally, Chikungunya and Zika cases, as well as the \textit{Aedes aegypti} infestation index, were not used for the models predicting cases in 2016, 2017, and 2018, as these covariates lacked values before 2015. The tuning parameter $\delta$ was selected to minimize BIC in two training sets: 2012-2015 for models used to predict years 2016, 2017, and 2019, and 2015-2018 for models used to predict years 2019 and 2020, resulting in $\delta = 0.00001$ for both. The tuning parameter $f$ was fixed at $f = 2$ to allow for the estimation of more parameters in long contexts.

\begin{figure}[h!] 
\hspace{-1.5cm}\includegraphics[scale=0.55]{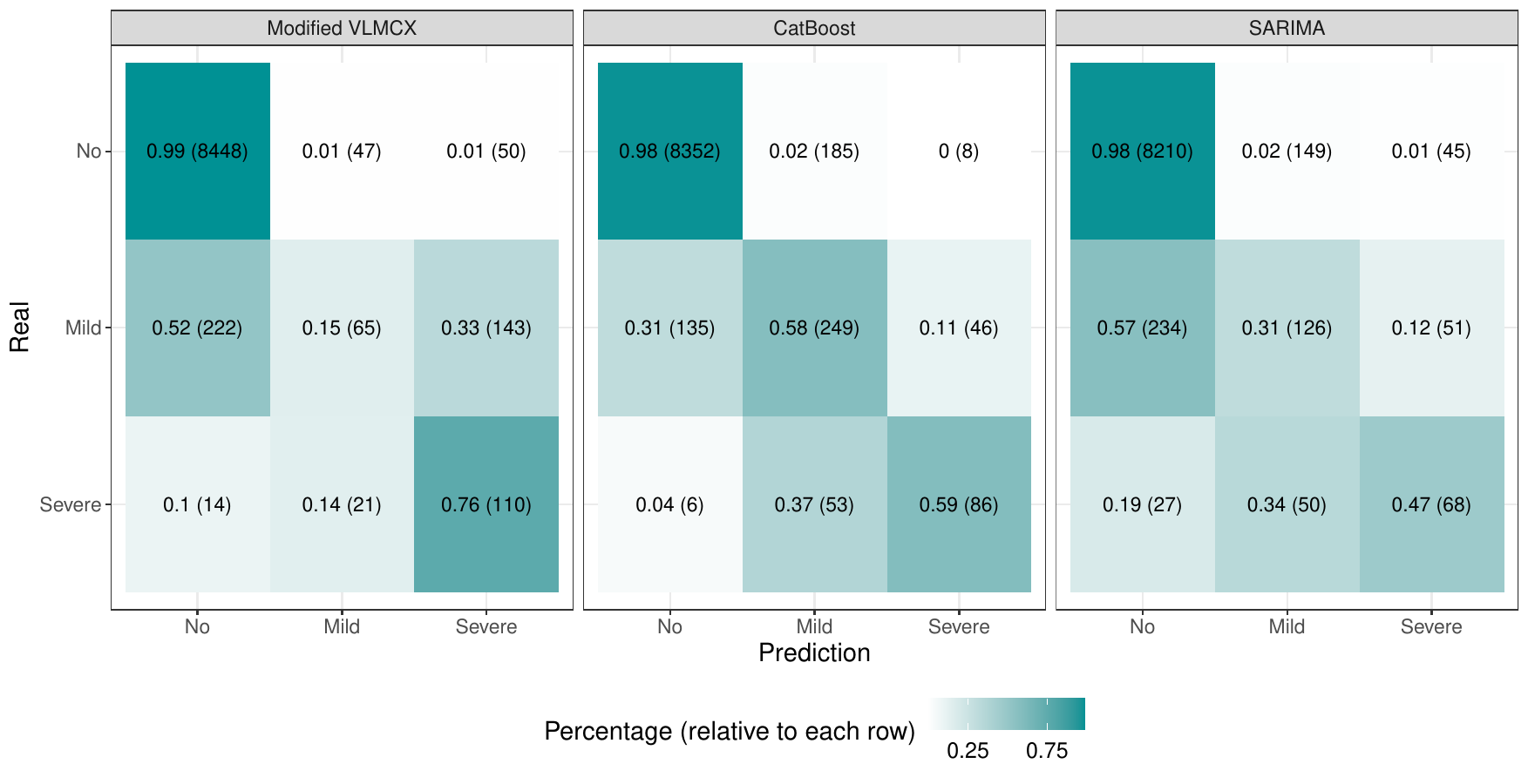}
\caption{\label{confusion_matrix_all}Confusion matrix for prediction results - Rio de Janeiro districts}
\end{figure}

\autoref{confusion_matrix_all} displays prediction results for our model and the ones fitted in \citet{aleixo2022predicting}. The predictions were made one month ahead, and all available observations from previous months were used to forecast each of the 12 months, as the model relies on covariates from previous months. Our model shows better prediction results for extreme categories, but it performs less well for intermediary categories. In terms of the worst-case scenario where the model predicts no outbreak while there is an outbreak, our model and SARIMA have similar results for this situation in the mild category. Specifically, our model predicted no outbreak for 52\% of the real values equal to mild, while this percentage is 57\% for the SARIMA model.

The current categorization of dengue cases may be influencing predictions, as it does not take into account the population size of each district and changes for each year being predicted. Additionally, since the categorization is based on the training set used for Catboost fitting, future years are being utilized to categorize cases, potentially introducing biases into the predictions.

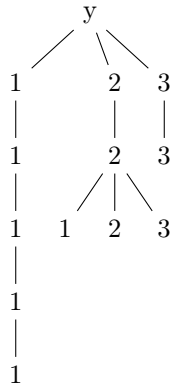
\begin{figure}[h!]
    \begin{center}
        \noindent
        $\begin{aligned}
          \begin{forest}
              [y 
                      [1
                          [1  
                              [1
                                [1
                                  [1]]]]]
                      [2
                          [2
                            [1]
                            [2]
                            [3]]]
                      [3 
                          [3]]]
          \end{forest}
          \end{aligned}$
        \noindent
    \end{center}
    \caption{\label{final_tree_rj_2016} Final tree for dengue incidence in Rio de Janeiro districts (fitted for years 2012-2015)}
\end{figure}

\begin{figure}[h!]
    \begin{center}
        \noindent
        $\begin{aligned}
          \begin{forest}
              [y 
                      [1]
                      [2
                          [1]
                          [2]
                          [3]]
                      [3 
                          [1]]]
          \end{forest}
          \end{aligned}$
    \end{center}
    \caption{\label{final_tree_rj_2017} Final tree for dengue incidence in Rio de Janeiro districts (fitted for years 2013-2016)}
\end{figure}
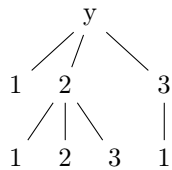

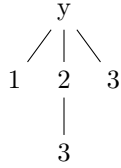
\begin{figure}[h!]
    \begin{center}
        \noindent
        $\begin{aligned}
          \begin{forest}
              [y 
                      [1]
                      [2
                          [3]]
                      [3]]
          \end{forest}
          \end{aligned}$
    \end{center}
    \caption{\label{final_tree_rj_2018} Final tree for dengue incidence in Rio de Janeiro districts (fitted for years 2014-2017)}
\end{figure}

\begin{figure}[h!]
    \begin{center}
        \noindent
        $\begin{aligned}
          \begin{forest}
              [y 
                      [1
                        [1
                          [1]]]
                      [2
                          [3]]
                      [3]]
          \end{forest}
          \end{aligned}$
    \end{center}
    \caption{\label{final_tree_rj_2019} Final tree for dengue incidence in Rio de Janeiro districts (fitted for years 2015-2018)}
\end{figure}
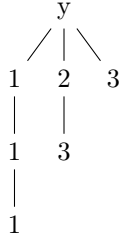

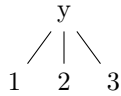
\begin{figure}[h!]
    \begin{center}
        \noindent
        $\begin{aligned}
          \begin{forest}
              [y 
                      [1]
                      [2]
                      [3]]
          \end{forest}
          \end{aligned}$
    \end{center}
    \caption{\label{final_tree_rj_2020} Final tree for dengue incidence in Rio de Janeiro districts (fitted for years 2016-2019)}
\end{figure}

Figures \ref{final_tree_rj_2016}--\ref{final_tree_rj_2020} display the estimated beta-contexts for each year, while Tables \ref{final_tree_rj_2016_estimates}--\ref{final_tree_rj_2020_estimates} provide the corresponding parameter estimates. The estimates for the time-varying exogenous covariates ($\hat{\boldsymbol{\beta}}$) follow this order: total precipitation, mean temperature, mean air humidity, the sum of dengue cases in neighboring districts, number of Zika cases, number of Chikungunya cases, and the \textit{Aedes aegypti} infestation index. For the time-invariant exogenous covariates, the order of the estimates is: demographic density, number of health facilities, and the sum of dengue cases in the past.

It is noticeable that the covariates with the most influence on future dengue cases are mean temperature (\degree{C}), mean air humidity (\%), total precipitation (mm), \textit{Aedes aegypti} infestation index, and the sum of dengue cases in the past (normalized to the 0 to 1 range). Additionally, in general, only two previous months are needed to predict the next one. Larger contexts occur when there is a history of months with low incidence (category 1). These findings align with those presented in \citet{aleixo2022predicting}, where the most relevant features for the CatBoost model were the number of cases in the previous month (cases $m-1$), followed by the dengue prevalence of the district, precipitation, neighbor cases, and temperature.

One advantage of our model over CatBoost (presented by \citet{aleixo2022predicting}) is that, while CatBoost allows the evaluation of the importance of variables in model predictions, our model provides the exact values of parameter estimates. This enables us to precisely understand how each covariate impacts dengue transmission in various scenarios of past dengue incidence. Furthermore, our model allows flexibility in examining the variable length of the past, providing precise insights into how past events influence future dengue transmission.

\begin{table}[h!]
\centering
\caption{Estimated parameters for dengue incidence in Rio de Janeiro districts (fitted for years 2012-2015)}
\label{final_tree_rj_2016_estimates}
\resizebox{\textwidth}{!}{%
\begin{tabular}{|c|c|c|c|}
\hline
\textbf{\begin{tabular}[c]{@{}c@{}}Contexts/\\ Estimators\end{tabular}} &
  \textbf{$\hat{\boldsymbol{\alpha}}$} &
  \textbf{$\hat{\boldsymbol{\beta}}$} &
  \textbf{$\hat{\boldsymbol{\gamma}}$} \\ \hline
\multirow{2}{*}{\textbf{1*}} &
  $\hat{\boldsymbol{\alpha}}^{1*}_{2} = -3.08$ &
  \multirow{2}{*}{} &
  \multirow{2}{*}{} \\
 &
  $\hat{\boldsymbol{\alpha}}^{1*}_{3} = -5.22$ &
   &
   \\ \hline
\multirow{2}{*}{\textbf{11*}} &
  $\hat{\boldsymbol{\alpha}}^{11*}_{2} = -3.94$ &
  \multirow{2}{*}{} &
  \multirow{2}{*}{} \\
 &
  $\hat{\boldsymbol{\alpha}}^{11*}_{3} = -37.32$ &
   &
   \\ \hline
\multirow{2}{*}{\textbf{111*}} &
  $\hat{\boldsymbol{\alpha}}^{111*}_{2} = -3.78$ &
  \multirow{2}{*}{} &
  \multirow{2}{*}{} \\
 &
  $\hat{\boldsymbol{\alpha}}^{111*}_{3} = -36.21$ &
   &
   \\ \hline
\multirow{2}{*}{\textbf{1111*}} &
  $\hat{\boldsymbol{\alpha}}^{1111*}_{2} = -3.72$ &
  \multirow{2}{*}{} &
  \multirow{2}{*}{} \\
 &
  $\hat{\boldsymbol{\alpha}}^{1111*}_{3} =  -5.8$ &
   &
   \\ \hline
\multirow{2}{*}{\textbf{11111}} &
  $\hat{\boldsymbol{\alpha}}^{11111}_{2} = -3.37$ &
  \multirow{2}{*}{} &
  $\hat{\boldsymbol{\gamma}}^{11111}_{2} = (0.01, 0, 3.37)'$ \\
 &
  $\hat{\boldsymbol{\alpha}}^{11111}_{3} = -5.64$ &
   &
  $\hat{\boldsymbol{\gamma}}^{11111}_{3} = (-0.04, 0, 5.42)'$ \\ \hline
\multirow{2}{*}{\textbf{2*}} &
  $\hat{\boldsymbol{\alpha}}^{2*}_{2} = -18.11$ &
  $\hat{\boldsymbol{\beta}}^{2*}_{2,1} = (0.11, 0.67, 0.01, 0)'$ &
  $\hat{\boldsymbol{\gamma}}^{2*}_{2} = (0.01, -0.01, 4.4)'$ \\
 &
  $\hat{\boldsymbol{\alpha}}^{2*}_{3} = -59.74$ &
  $\hat{\boldsymbol{\beta}}^{2*}_{3,1} = (0.11, 1.78, 0.18, 0)'$ &
  $\hat{\boldsymbol{\gamma}}^{2*}_{3} = (0.02, 0.01, 8.22)'$ \\ \hline
\multirow{6}{*}{\textbf{221}} &
  \multirow{3}{*}{$\hat{\boldsymbol{\alpha}}^{221}_{2} = -106.82$} &
  $\hat{\boldsymbol{\beta}}^{221}_{2,1} = (3.14, 9.62, 5.27, 0)'$ &
  \multirow{3}{*}{$\hat{\boldsymbol{\gamma}}^{221}_{2} = (-0.01, 0.03, 145.09)'$} \\
 &
   &
  $\hat{\boldsymbol{\beta}}^{221}_{2,2} = (2.74, -7.98, -0.71, -0.03)'$ &
   \\
 &
   &
  $\hat{\boldsymbol{\beta}}^{221}_{2,3} = (2.03, -6.7, -2.22, 0.06)'$ &
   \\
 &
  \multirow{3}{*}{$\hat{\boldsymbol{\alpha}}^{221}_{3} = -164.7$} &
  $\hat{\boldsymbol{\beta}}^{221}_{3,1} = (2.91, 10.2, 4.61, 0)'$ &
  \multirow{3}{*}{$\hat{\boldsymbol{\gamma}}^{221}_{3} = (0.02, 0.06, 138.42)'$} \\
 &
   &
  $\hat{\boldsymbol{\beta}}^{221}_{3,2} = (2.74, -6.81, -0.44, -0.02)'$ &
   \\
 &
   &
  $\hat{\boldsymbol{\beta}}{221}_{3,3} = (2.02, -6.57, -1.65,  0.03)'$ &
   \\ \hline
\multirow{2}{*}{\textbf{222}} &
  $\hat{\boldsymbol{\alpha}}^{222}_{2} = -1.9$ &
  \multirow{2}{*}{} &
  $\hat{\boldsymbol{\gamma}}^{222}_{2} = (0.05, 0.01, 2.74)'$ \\
 &
  $\hat{\boldsymbol{\alpha}}^{222}_{3} = -2.84$ &
   &
  $\hat{\boldsymbol{\gamma}}^{222}_{3} = (0.07, 0.02, 1.46)'$ \\ \hline
\multirow{2}{*}{\textbf{223}} &
  $\hat{\boldsymbol{\alpha}}^{223}_{2} = -1.73$ &
  \multirow{2}{*}{} &
  \multirow{2}{*}{} \\
 &
  $\hat{\boldsymbol{\alpha}}^{223}_{3} =  -11.42$ &
   &
   \\ \hline
\multirow{2}{*}{\textbf{3*}} &
  $\hat{\boldsymbol{\alpha}}^{3*}_{2} = 2.56$ &
  \multirow{2}{*}{} &
  \multirow{2}{*}{} \\
 &
  $\hat{\boldsymbol{\alpha}}^{3*}_{3} = 4.65$ &
   &
   \\ \hline
\multirow{4}{*}{\textbf{33*}} &
  \multirow{2}{*}{$\hat{\boldsymbol{\alpha}}^{33*}_{2} = -9.52$} &
  $\hat{\boldsymbol{\beta}}^{33*}_{2,1} = (-0.21, -0.48, 0.01, 0)'$ &
  \multirow{2}{*}{$\hat{\boldsymbol{\gamma}}^{33*}_{2} = (0, 0, 0.05)'$} \\
 &
   &
  $\hat{\boldsymbol{\beta}}^{33*}_{2,2} = (0.03, 0.38, 0.17, 0)'$ &
   \\
 &
  \multirow{2}{*}{$\hat{\boldsymbol{\alpha}}^{33*}_{3} = -52.91$} &
  $\hat{\boldsymbol{\beta}}^{33*}_{3,1} = (-0.11, 0.37, 0.2, 0)'$ &
  \multirow{2}{*}{$\hat{\boldsymbol{\gamma}}^{33*}_{3} = (0.03, 0, 7.12)'$} \\
 &
   &
  $\hat{\boldsymbol{\beta}}^{33*}_{3,2} = (-0.03, 1.08, 0.04, 0)'$ &
   \\ \hline
   \multicolumn{4}{c}{%
            \begin{minipage}{17cm}%
                \footnotesize
                \vspace{1mm} \textbf{Note:} $\hat{\alpha}_{j}^{u}$ represents the estimated intercept,  $\hat{\boldsymbol{\gamma}}_{j}^{u}$ the estimated coefficients for the time-invariant covariates (with coordinates corresponding to demographic density, number of health facilities, and the sum of dengue cases in the past, respectively) and, $\hat{\boldsymbol{\beta}}_{j,t}^{u}$ represents the estimated coefficients for time-dependent covariates (with coordinates corresponding to total precipitation, mean temperature, mean air humidity, the sum of dengue cases in neighboring districts, number of Zika cases, number of Chikungunya cases, and the \textit{Aedes aegypti} infestation index, respectively) that influence the transition probability from context $u$ to state $j$ at time $t$.
            \end{minipage}%
        }
\end{tabular}}%
\end{table}

\begin{table}[h!]
\centering
\caption{Estimated parameters for dengue incidence in Rio de Janeiro districts (fitted for years 2013-2016)}
\label{final_tree_rj_2017_estimates}
\resizebox{\textwidth}{!}{%
\begin{tabular}{|c|c|c|c|}
\hline
\textbf{\begin{tabular}[c]{@{}c@{}}Contexts/\\ Estimators\end{tabular}} &
  \textbf{$\hat{\boldsymbol{\alpha}}$} &
  \textbf{$\hat{\boldsymbol{\beta}}$} &
  \textbf{$\hat{\boldsymbol{\gamma}}$} \\ \hline
\multirow{2}{*}{\textbf{1*}} &
  $\hat{\boldsymbol{\alpha}}^{1*}_{2} = -36.31$ &
  $\hat{\boldsymbol{\beta}}^{1*}_{2,1} = (0.03, 0.76, 0.17, 0.01)'$ &
  $\hat{\boldsymbol{\gamma}}^{1*}_{2} = (0.02, 0, 4.46)'$ \\
 &
  $\hat{\boldsymbol{\alpha}}^{1*}_{3} = -49.46$ &
  $\hat{\boldsymbol{\beta}}^{1*}_{3,1} = (0, 0.9, 0.26, 0.01)'$ &
  $\hat{\boldsymbol{\gamma}}^{1*}_{3} = (0.02, -0.02, 7.57)'$ \\ \hline
\multirow{4}{*}{\textbf{21}} &
  \multirow{2}{*}{$\hat{\boldsymbol{\alpha}}^{21}_{2} = 3.48$} &
  $\hat{\boldsymbol{\beta}}^{21}_{2,1} = (-0.05, 1.44, 0.68, 0.01)'$ &
  \multirow{2}{*}{$\hat{\boldsymbol{\gamma}}^{21}_{2} = (0.01, 0, 2.59)'$} \\
 &
   &
  $\hat{\boldsymbol{\beta}}^{21}_{2,2} = (-0.01, -1.64, -0.63, -0.02)'$ &
   \\
 &
  \multirow{2}{*}{$\hat{\boldsymbol{\alpha}}^{21}_{3} = 8.81$} &
  $\hat{\boldsymbol{\beta}}^{21}_{3,1} = (-0.13, 2.44, 0.89, 0.01)'$ &
  \multirow{2}{*}{$\hat{\boldsymbol{\gamma}}^{21}_{3} = (0.03, 0, 3.76)'$} \\
 &
   &
  $\hat{\boldsymbol{\beta}}^{21}_{3,2} = (-0.04, -2.68, -0.89, -0.02)'$ &
   \\ \hline
\multirow{2}{*}{\textbf{22}} &
  $\hat{\boldsymbol{\alpha}}^{22}_{2} = -12.59$ &
  $\hat{\boldsymbol{\beta}}^{22}_{2,1} = (0.14, 0.35, 0.02, 0)'$ &
  $\hat{\boldsymbol{\gamma}}^{22}_{2} = (0, 0, 4.93)'$ \\ 
 &
  $\hat{\boldsymbol{\alpha}}^{22}_{3} = -91.4$ &
  $\hat{\boldsymbol{\beta}}^{22}_{3,1} = (0.05, 2.39, 0.37, 0)'$ &
  $\hat{\boldsymbol{\gamma}}^{22}_{3} = (0.02, -0.03, 5.76)'$ \\ \hline
\multirow{2}{*}{\textbf{23}} &
  $\hat{\boldsymbol{\alpha}}^{23}_{2} = -2.51$ &
  \multirow{2}{*}{} &
  \multirow{2}{*}{} \\
 &
  $\hat{\boldsymbol{\alpha}}^{23}_{3} = -12.08$ &
   &
   \\ \hline
\multirow{2}{*}{\textbf{3*}} &
  $\hat{\boldsymbol{\alpha}}^{3*}_{2} = -5.83$ &
  $\hat{\boldsymbol{\beta}}{3*}_{2,1} = (0.01, 0.09, 0.04, 0)'$ &
  $\hat{\boldsymbol{\gamma}}^{3*}_{2} = (0.04, -0.01, 2.54)'$ \\
 &
  $\hat{\boldsymbol{\alpha}}^{3*}_{3} = -38.49$ &
  $\hat{\boldsymbol{\beta}}^{3*}_{3,1} = (0.14, 0.89, 0.18, 0)'$ &
  $\hat{\boldsymbol{\gamma}}^{3*}_{3} = (0.05, 0, 6.72)'$ \\ \hline
\multirow{2}{*}{\textbf{31}} &
  $\hat{\boldsymbol{\alpha}}^{31}_{2} = 6.92$ &
  \multirow{2}{*}{} &
  \multirow{2}{*}{} \\
 &
  $\hat{\boldsymbol{\alpha}}^{31}_{3} = 8.94$ &
   &
   \\ \hline
   \multicolumn{4}{c}{%
            \begin{minipage}{17cm}%
                \footnotesize
                \vspace{1mm} \textbf{Note:} $\hat{\alpha}_{j}^{u}$ represents the estimated intercept,  $\hat{\boldsymbol{\gamma}}_{j}^{u}$ the estimated coefficients for the time-invariant covariates (with coordinates corresponding to demographic density, number of health facilities, and the sum of dengue cases in the past, respectively) and, $\hat{\boldsymbol{\beta}}_{j,t}^{u}$ represents the estimated coefficients for time-dependent covariates (with coordinates corresponding to total precipitation, mean temperature, mean air humidity, the sum of dengue cases in neighboring districts, number of Zika cases, number of Chikungunya cases, and the \textit{Aedes aegypti} infestation index, respectively) that influence the transition probability from context $u$ to state $j$ at time $t$.
            \end{minipage}%
        }
\end{tabular}%
}
\end{table}

\begin{table}[h!]
\centering
\caption{Estimated parameters for dengue incidence in Rio de Janeiro districts (fitted for years 2014-2017)}
\label{final_tree_rj_2018_estimates}
\resizebox{\textwidth}{!}{%
\begin{tabular}{|c|c|c|c|}
\hline
\textbf{\begin{tabular}[c]{@{}c@{}}Contexts/\\ Estimators\end{tabular}} &
  \textbf{$\hat{\boldsymbol{\alpha}}$} &
  \textbf{$\hat{\boldsymbol{\beta}}$} &
  \textbf{$\hat{\boldsymbol{\gamma}}$} \\ \hline
\multirow{2}{*}{\textbf{1}}  & $\hat{\boldsymbol{\alpha}}^{1}_{2} = -3.92$   & \multirow{2}{*}{} & \multirow{2}{*}{} \\
                             & $\hat{\boldsymbol{\alpha}}^{1}_{3} = -7.23$   &                   &                   \\ \hline
\multirow{2}{*}{\textbf{2*}} &
  $\hat{\boldsymbol{\alpha}}^{2*}_{2} = -19.78$ &
  $\hat{\boldsymbol{\beta}}^{2*}_{2,1} = (0.07, 0.23, 0.18, 0)'$ &
  $\hat{\boldsymbol{\gamma}}^{2*}_{2} = (0.01, 0, 2.27)'$ \\
 &
  $\hat{\boldsymbol{\alpha}}^{2*}_{3} = -32.16$ &
  $\hat{\boldsymbol{\beta}}^{2*}_{3,1} = (0.18, 0.36, 0.24, 0)'$ &
  $\hat{\boldsymbol{\gamma}}^{2*}_{3} = (0.03, -0.03, 6.08)'$ \\ \hline
\multirow{2}{*}{\textbf{23}} & $\hat{\boldsymbol{\alpha}}^{23}_{2} = -2.04$  & \multirow{2}{*}{} & \multirow{2}{*}{} \\
                             & $\hat{\boldsymbol{\alpha}}^{23}_{3} = -11.29$ &                   &                   \\ \hline
\multirow{2}{*}{\textbf{3}}  & $\hat{\boldsymbol{\alpha}}^{3}_{2} = 1.87$    & \multirow{2}{*}{} & \multirow{2}{*}{} \\
                             & $\hat{\boldsymbol{\alpha}}^{3}_{3} = 1.98$    &                   &                   \\ \hline
                             \multicolumn{4}{c}{%
            \begin{minipage}{17cm}%
                \footnotesize
                \vspace{1mm} \textbf{Note:} $\hat{\alpha}_{j}^{u}$ represents the estimated intercept,  $\hat{\boldsymbol{\gamma}}_{j}^{u}$ the estimated coefficients for the time-invariant covariates (with coordinates corresponding to demographic density, number of health facilities, and the sum of dengue cases in the past, respectively) and, $\hat{\boldsymbol{\beta}}_{j,t}^{u}$ represents the estimated coefficients for time-dependent covariates (with coordinates corresponding to total precipitation, mean temperature, mean air humidity, the sum of dengue cases in neighboring districts, number of Zika cases, number of Chikungunya cases, and the \textit{Aedes aegypti} infestation index, respectively) that influence the transition probability from context $u$ to state $j$ at time $t$.
            \end{minipage}%
        }
\end{tabular}%
}
\end{table}

\begin{table}[h!]
\centering
\caption{Estimated parameters for dengue incidence in Rio de Janeiro districts (fitted for years 2015-2018)}
\label{final_tree_rj_2019_estimates}
\resizebox{\textwidth}{!}{%
\begin{tabular}{|c|c|c|c|}
\hline
\textbf{\begin{tabular}[c]{@{}c@{}}Contexts/\\ Estimators\end{tabular}} &
  \textbf{$\hat{\boldsymbol{\alpha}}$} &
  \textbf{$\hat{\boldsymbol{\beta}}$} &
  \textbf{$\hat{\boldsymbol{\gamma}}$} \\ \hline
\multirow{2}{*}{\textbf{1*}} &
  $\hat{\boldsymbol{\alpha}}^{1*}_{2} = -2.92$ &
  \multirow{2}{*}{} &
  \multirow{2}{*}{} \\
 &
  $\hat{\boldsymbol{\alpha}}^{1*}_{3} = -5.23$ &
   &
   \\ \hline
\multirow{2}{*}{\textbf{11*}} &
  $\hat{\boldsymbol{\alpha}}^{11*}_{2} = -3.4$ &
  \multirow{2}{*}{} &
  \multirow{2}{*}{} \\
 &
  $\hat{\boldsymbol{\alpha}}^{11*}_{3} = -17.44$ &
   &
   \\ \hline
\multirow{2}{*}{\textbf{111}} &
  $\hat{\boldsymbol{\alpha}}^{111}_{2} = -4$ &
  \multirow{2}{*}{} &
  \multirow{2}{*}{} \\
 &
  $\hat{\boldsymbol{\alpha}}^{111}_{3} = -6.92$ &
   &
   \\ \hline
\multirow{2}{*}{\textbf{2*}} &
  $\hat{\boldsymbol{\alpha}}^{2*}_{2} = -19.94$ &
  $\hat{\boldsymbol{\beta}}^{2*}_{2,1} = (0.03, 0.26, 0.12, 0, 0, -0.01, 4.56)'$ &
  $\hat{\boldsymbol{\gamma}}^{2*}_{2} = (0.01, 0, 2.56)'$ \\
 &
  $\hat{\boldsymbol{\alpha}}^{2*}_{3} = -53.96$ &
  $\hat{\boldsymbol{\beta}}^{2*}_{3,1} = (0.12, 0.67, 0.3, 0, 0, 0.01, 9.58)'$ &
  $\hat{\boldsymbol{\gamma}}^{2*}_{3} = (0.01, -0.03, 6.81)'$ \\ \hline
\multirow{2}{*}{\textbf{23}} &
  $\hat{\boldsymbol{\alpha}}^{23}_{2} = -0.81$ &
  \multirow{2}{*}{} &
  \multirow{2}{*}{} \\
 &
  $\hat{\boldsymbol{\alpha}}^{23}_{3} = -13.02$ &
   &
   \\ \hline
\multirow{2}{*}{\textbf{3}} &
  $\hat{\boldsymbol{\alpha}}^{3}_{2} = 0.95$ &
  \multirow{2}{*}{} &
  $\hat{\boldsymbol{\gamma}}^{3}_{2} = (0.02, 0.05, -1.64)'$ \\
 &
  $\hat{\boldsymbol{\alpha}}^{3}_{3} = 0.57$ &
   &
  $\hat{\boldsymbol{\gamma}}^{3}_{3} = (0.02, 0.04, 1.03)'$ \\ \hline
  \multicolumn{4}{c}{%
            \begin{minipage}{17cm}%
                \footnotesize
                \vspace{1mm}\textbf{Note:} $\hat{\alpha}_{j}^{u}$ represents the estimated intercept,  $\hat{\boldsymbol{\gamma}}_{j}^{u}$ the estimated coefficients for the time-invariant covariates (with coordinates corresponding to demographic density, number of health facilities, and the sum of dengue cases in the past, respectively) and, $\hat{\boldsymbol{\beta}}_{j,t}^{u}$ represents the estimated coefficients for time-dependent covariates (with coordinates corresponding to total precipitation, mean temperature, mean air humidity, the sum of dengue cases in neighboring districts, number of Zika cases, number of Chikungunya cases, and the \textit{Aedes aegypti} infestation index, respectively) that influence the transition probability from context $u$ to state $j$ at time $t$.
            \end{minipage}%
        }
\end{tabular}%
}
\end{table}

\begin{table}[h!]
\centering
\caption{Estimated parameters for dengue incidence in Rio de Janeiro districts (fitted for years 2016-2019)}
\label{final_tree_rj_2020_estimates}
\resizebox{\textwidth}{!}{%
\begin{tabular}{|c|c|c|c|}
\hline
\textbf{\begin{tabular}[c]{@{}c@{}}Contexts/\\ Estimators\end{tabular}} &
  \textbf{$\hat{\boldsymbol{\alpha}}$} &
  \textbf{$\hat{\boldsymbol{\beta}}$} &
  \textbf{$\hat{\boldsymbol{\gamma}}$} \\ \hline
\multirow{2}{*}{\textbf{1}} & $\hat{\boldsymbol{\alpha}}^{1}_{2} = -3.92$ & \multirow{2}{*}{} & \multirow{2}{*}{} \\
                            & $\hat{\boldsymbol{\alpha}}^{1}_{3} = -7.44$ &                   &                   \\ \hline
\multirow{2}{*}{\textbf{2}} &
  $\hat{\boldsymbol{\alpha}}^{2}_{2} = -39.13$ &
  $\hat{\boldsymbol{\beta}}^{2}_{2,1} = (0.08, 0.6, 0.34, 0, 0, 0, -4.04)'$ &
  $\hat{\boldsymbol{\gamma}}^{2}_{2} = (0.02, -0.01, 3.64)'$ \\
 &
  $\hat{\boldsymbol{\alpha}}^{2}_{3} = -90.22$ &
  $\hat{\boldsymbol{\beta}}^{2}_{3,1} = (0.17, 1.64, 0.55, 0.01, 0, 0.01, -1.59)'$ &
  $\hat{\boldsymbol{\gamma}}^{2}_{3} = (0.05, -0.05, 10.35)'$ \\ \hline
\multirow{2}{*}{\textbf{3}} & $\hat{\boldsymbol{\alpha}}^{3}_{2} = 1.73$  & \multirow{2}{*}{} & \multirow{2}{*}{} \\
                            & $\hat{\boldsymbol{\alpha}}^{3}_{3} = 1.76$  &                   &                   \\ \hline
                            \multicolumn{4}{c}{%
            \begin{minipage}{17cm}%
                \footnotesize
                \vspace{1mm}\textbf{Note:} $\hat{\alpha}_{j}^{u}$ represents the estimated intercept,  $\hat{\boldsymbol{\gamma}}_{j}^{u}$ the estimated coefficients for the time-invariant covariates (with coordinates corresponding to demographic density, number of health facilities, and the sum of dengue cases in the past, respectively) and, $\hat{\boldsymbol{\beta}}_{j,t}^{u}$ represents the estimated coefficients for time-dependent covariates (with coordinates corresponding to total precipitation, mean temperature, mean air humidity, the sum of dengue cases in neighboring districts, number of Zika cases, number of Chikungunya cases, and the \textit{Aedes aegypti} infestation index, respectively) that influence the transition probability from context $u$ to state $j$ at time $t$.
            \end{minipage}%
        }
\end{tabular}%
}
\end{table}

\section{Conclusion and future work}

This study has successfully expanded the capabilities of the beta-context algorithm, integrating both time-dependent and time-invariant exogenous covariates from multiple independent sources and addressing challenges related to rare events and limited data, providing a more versatile modeling framework.  Although the algorithm is similar to the one proposed in \citet{zanin2022variable}, several modifications are worth emphasizing. 

\begin{enumerate}
    \item \textbf{Time-invariant exogenous covariates:} Time-fixed parameters ($\gamma_j^u$) for time-invariant exogenous covariates were introduced for each terminal node. Here, $u$ denotes a terminal node, and $j = 1, ..., p$ represents the transition to state $j$. Note that during the tree-pruning process, the time-fixed parameter ($\gamma_j^u$) is pruned simultaneously with the intercept ($\alpha_j^u$) when nodes are lumped together. \\
    
    \item \textbf{Observations from various independent sources:} This aspect involves integrating data from several independent sources, allowing for a comprehensive understanding of the process behavior across multiple situations, such as different geographic locations. In this context, time-invariant exogenous covariates may vary for each source, providing a more nuanced representation of fixed conditions. Despite these variations, model parameters are assumed to be the same across sources, enhancing the predictive capabilities of the model. The current model does not incorporate a spatial-dependent structure, assuming sources are independent. \\
    
    \item \textbf{Rare events:} To ensure the model's robustness in handling rare events, particularly in scenarios with finite observations, we have implemented several strategies:
    
    \begin{itemize}
        \item \textbf{Minimum observation requirement:} Each category within every terminal node is subjected to a minimum observation threshold, ensuring a sufficient number of occurrences for reliable parameter estimation.
        \item \textbf{Partial parameter estimation:} Terminal nodes have the flexibility to possess incomplete sets of parameter estimates. If the available observations are not enough for estimating all parameters, only a subset is computed. The intercept remains a consistently estimated parameter.
        \item \textbf{Intercept Estimation with Neural Network:} In cases where a specific event never occurs for a terminal node, the intercept can be estimated using a single-hidden-layer neural network \citep{venables2002modern}. 
        \item \textbf{Exact Tests for Rare Events:} The model incorporates exact tests, such as the Fisher and Cochran-Mantel-Haenszel tests, which are particularly useful in situations involving rare events.
    \end{itemize}
    
    These measures collectively increase the model's adaptability to scenarios characterized by infrequent events and a limited number of observations.
    
    An alternative approach considered was to set a fixed probability for the rare event while estimating all parameters for the other possible categories. However, challenges arose in determining how to calculate estimates with the bound on the fixed probability. Additionally, complications emerged when the rare event served as the baseline for other branches. \\
    
    \item \textbf{Terminal node completeness requirement for initial tree:} Although the initial tree $\tau^{(0)}$ may lack completeness, a crucial constraint is imposed on all terminal nodes ‚Äî they must have all possible children. This constraint is imposed because, in the presence of children, we refrain from estimating parameters for the parent node. Consequently, without this constraint, there might be insufficient occurrences for a child node to be represented in the tree, leading to a lack of parameter estimates for that particular node.
\end{enumerate}

Simulations conducted without time-invariant exogenous covariates and with only one source demonstrated that the modified algorithm outperforms the original version, particularly in situations with limited data. The modified version provided recovery of more accurate tree structures and covariate vectors. For simulations with time-invariant exogenous covariates and multiple independent sources, improvements were observed in the percentage of identical trees and exact covariate vectors with increasing sample size. These results can be seen in the online Supplementary Material. 

Our motivation comes from analyzing a real dataset focused on monthly dengue cases across multiple municipalities in Brazil. The inclusion of time-dependent covariates - temperature and precipitation levels, along with time-invariant covariates - poverty rates and urban population percentages, allowed us to investigate the complex dynamics of dengue transmission. The model demonstrated satisfactory performance in predicting dengue incidence, especially for high-incidence cases, showing its potential for early detection and proactive management of outbreaks. 

Looking forward, our future work includes incorporating additional model possibilities, including non-parametric methods and spatial correlation to account for non-independent multiple sources. Besides that, we aim to develop mechanisms to: handle missing values in exogenous covariates, address the challenge of impossible contexts, test the significance of time-invariant exogenous covariates, and enable estimation of exogenous covariates parameters when not all categories/states for a specific context are available. Concerning dengue outbreak prediction, we aim to conduct a more detailed study on how covariates influence dengue cases, exploring non-linear effects and interactions between factors. The overarching objective is to continually enhance the model's flexibility, accuracy, and applicability across diverse settings. Additionally, plans involve the publication of the algorithm implementation as a package on The Comprehensive R Archive Network (CRAN), further contributing to the broader scientific community. \\

\noindent{\bf Acknowledgments: } MGR was financed in part by the Coordenação de Aperfeiçoamento de Pessoal de Nível Superior – Brasil (CAPES) – Finance Code 001. NLG was financed by FAPESP grants 2019/04535-2 and 2023/13453-5, CNPq grants 302598/2014-6, 442012/2014-4 and 304148/2020-2. \\

\bibliographystyle{apalike}
\bibliography{dengue_arxiv.bib}

\end{document}